\documentclass{emulateapj}
\usepackage{graphicx}
\usepackage{epstopdf}

\shorttitle{Herschel-PACS observations of OH at 119$\mu$m}
\shortauthors{Spoon et al.}

\begin{document}


\title{Diagnostics of AGN-driven molecular outflows in ULIRGs 
from Herschel-PACS observations of OH at 119$\mu$m}


\author{H.W.W. Spoon\altaffilmark{1}}
\author{D. Farrah\altaffilmark{2}}
\author{V. Lebouteiller\altaffilmark{3,1}}
\author{E. Gonz\'alez-Alfonso\altaffilmark{4}}
\author{J. Bernard-Salas\altaffilmark{5}}
\author{T. Urrutia\altaffilmark{6}}
\author{D. Rigopoulou\altaffilmark{7,8}}
\author{M.S. Westmoquette\altaffilmark{9}}
\author{H.A. Smith\altaffilmark{10}}
\author{J. Afonso\altaffilmark{11,12}}
\author{C. Pearson\altaffilmark{13,14}}
\author{D. Cormier\altaffilmark{15}}
\author{A. Efstathiou\altaffilmark{16}}
\author{C. Borys\altaffilmark{17}}
\author{A. Verma\altaffilmark{7}}
\author{M. Etxaluze\altaffilmark{18}}
\author{D.L. Clements\altaffilmark{19}}

\altaffiltext{1}{Cornell University, CRSR, Space Sciences Building, Ithaca, NY 14853, USA; spoon@isc.astro.cornell.edu} 
\altaffiltext{2}{Department of Physics, Virginia Tech, Blacksburg, VA  24061, USA} 
\altaffiltext{3}{CEA-Saclay, DSM/IRFU/SAp, F-91191 Gif-sur-Yvette,  France} 
\altaffiltext{4}{Universidad de Alcal\'a, Departamento de F\'{\i}sica y Matem\'aticas, 
Campus Universitario, E-28871 Alcal\'a de Henares, Madrid, Spain} 
\altaffiltext{5}{Open University, Department of Physical Sciences, Milton Keynes, MK7 6AA, UK} 
\altaffiltext{6}{Leibniz Institut f\"{u}r Astrophysik, Potsdam, An der Sternwarte 16, 14482, Potsdam, Germany} 
\altaffiltext{7}{Department of Physics, Denys Wilkinson Building, Keble Road, Oxford, OX1 3RH, UK} 
\altaffiltext{8}{RAL Space, Science \& Technology Facilities Council, Rutherford Appleton Laboratory, Didcot, OX11 0QX, UK} 

\altaffiltext{9}{European Southern Observatory, Karl-Schwarzschild-Str. 2, 85748 Garching bei M\"unchen, Germany} 
\altaffiltext{10}{Harvard-Smithsonian Center for Astrophysics, 60 Garden Street, Cambridge, MA 02138, USA} 
\altaffiltext{11}{Centro de Astronomia e Astrof\'{\i}sica da Universidade de Lisboa, Observat\'{o}rio Astron\'{o}mico de Lisboa, Tapada da Ajuda, 1349-018 Lisbon, Portugal} 
\altaffiltext{12}{Department of Physics, Faculty of Sciences, University of Lisbon, Campo Grande, 1749-016 Lisbon, Portugal} 
\altaffiltext{13}{RAL Space, Rutherford Appleton Laboratory, Harwell, Oxford, OX11 0QX, U.K.} 
\altaffiltext{14}{Dept. of Physics \& Astronomy, The Open University, Milton Keynes, MK7 6AA, U.K.} 
\altaffiltext{15}{Institut f\"ur theoretische Astrophysik, Zentrum f\"ur Astronomie der Universit\"at Heidelberg, Albert-Ueberle Str. 2, D-69120 Heidelberg, Germany} 
\altaffiltext{16}{School of Sciences, European University Cyprus, Diogenes Street, Engomi, 1516 Nicosia, Cyprus} 
\altaffiltext{17}{Infrared Processing and Analysis Center, California Institute of Technology, MS 100-22, 
Pasadena, CA 91125, USA} 
\altaffiltext{18}{Departamento de Astrof\'isica. Centro de Astrobiolog\'ia. CSIC-INTA. Torrej\'on de Ardoz, 28850 Madrid, Spain} 
\altaffiltext{19}{Physics Department, Imperial College London, Prince Consort Road, London, SW7 2AZ, UK} 

\begin{abstract}
We report on our observations of the 79 and 119\,$\mu$m doublet
transitions of OH for 24 local (z$<$0.262) ULIRGs observed with 
Herschel-PACS as part of the Herschel ULIRG Survey (HERUS). 
Some OH119 profiles display a clear 
P-Cygni shape and therefore imply outflowing OH gas, other
profiles are predominantly in absorption or are completely 
in emission. We find that the relative strength of the OH 
emission component decreases as the silicate absorption increases. 
This locates the OH outflows inside the obscured nuclei.
The maximum outflow velocities for our sources
range from less than 100 to $\sim$2000\,km s$^{-1}$, with 15/24 
(10/24) sources showing OH absorption at velocities exceeding
700\,km s$^{-1}$ (1000\,km s$^{-1}$). Three sources show maximum 
OH outflow velocities exceeding that of Mrk231. 
Since outflow velocities above 500--700\,km s$^{-1}$ are 
thought to require an active galactic nucleus (AGN) to drive 
them, about 2/3 of our ULIRG sample may host AGN-driven molecular 
outflows.
This finding is supported by the correlation we find between 
the maximum OH outflow velocity and the IR-derived bolometric 
AGN luminosity. No such correlation is found with the IR-derived 
star formation rate. 
The highest outflow velocities are found among sources which are 
still deeply embedded. We speculate that the molecular outflows in 
these sources may be in an early phase of disrupting the nuclear 
dust veil before these sources evolve into less obscured AGN.
Four of our sources show high-velocity wings in their [C II] 
fine-structure line profiles implying neutral gas outflow 
masses of at least 2--4.5$\times$10$^8$ M$_{\odot}$.
\end{abstract}



\section{Introduction}

The degree to which Active Galactic Nuclei (AGN) affect their kiloparsec to
megaparsec scale surroundings is a key idea in understanding the
cosmological assembly history of galaxies. This idea, commonly termed 
`AGN feedback', can take several forms; we here focus on the specific 
case where a central supermassive black hole (SMBH) exerts a strong 
influence on short
($\sim10^7$ yr) timescales and $\sim$kpc spatial scales to quench star 
formation in its host galaxy. Such feedback is thought to occur when 
radiation from the accretion disk drives a wind of partially ionized 
gas into the host, which then empties the galaxy of fuel for star
formation by kinetically driving the ISM gas out of the galaxy, and/or 
heating the ISM so it cannot collapse to form stars. It is thought to 
be particularly effective in galaxy mergers, after the progenitor
nuclei have coalesced \citep{dimatteo05,debu12,choi12}. IR-luminous 
galaxy mergers, which often harbor both high rates of star formation 
and luminous AGN, are thus prime sites to study the properties and 
effects of AGN feedback. 

There is strong theoretical motivation for AGN feedback from models 
that trace the assembly of galaxies with redshift. Models without 
quasar mode feedback have great difficulty in reproducing the observed 
redshift evolution of the galaxy mass function \citep{somer01,benson03} 
whereas models with a prescription for AGN feedback show dramatic 
improvements in their consistency 
with observations \citep{croton06,somer08}. Moreover, simulations of 
individual galaxies suggest that quasar mode feedback has a profound 
influence \citep[e.g.][]{hopkins10}. Observational diagnostics of AGN 
feedback are however relatively sparse. It has been inferred indirectly 
from e.g. the discovery of winds with high kinetic fluxes in ULIRGs 
and QSOs \citep{moe09,rupke11}, and directly from e.g. the star formation 
properties of QSO host galaxies \citep{farrah12}. Though some recent 
claims for quasar driven suppression of star formation \citep{page12} 
are controversial \citep{harr12}. 

\begin{deluxetable*}{lllll}
\tablewidth{0pt}
\tablecaption{PACS observations log\label{tbl-0}}
\tablehead{
\colhead{Galaxy} & \colhead{Line} & \colhead{Obs Id} & \colhead{Obs Date} & \colhead{spaxel used\tablenotemark{a}}
}
\startdata
00188--0856 & OH119 \& OH79 & 1342237570 & 2012 Jan 16  & (3,3)\\
00397--1312 & OH79          & 1342238351 & 2012 Jan 25  & (3,3)\\
00397--1312 & OH119         & 1342238350 & 2012 Jan 25  & (3,3)\\
01003--2238 & OH119 \& OH79 & 1342238371 & 2012 Jan 28  & (3,3)\\
Mrk1014     & OH119 \& OH79 & 1342238723 & 2012 Feb 6  & (3,3)\\
03158+4227  & OH119 \& OH79 & 1342238963 & 2012 Feb 11 & (3,3)\\
03521+0028  & OH119 \& OH79 & 1342239746 & 2012 Feb 26 & (3,3)\\
06035--7102 & OH119 \& OH79 & 1342239479 & 2012 Feb 15 & (3,3)\\
06206--6315 & OH119 \& OH79 & 1342230961 & 2011 Oct 14  & (3,3)\\
07598+6508  & OH79\tablenotemark{b}      & 1342243534 & 2012 Mar 25 & (3,3)\\
07598+6508  & OH119\tablenotemark{c}     & 1342231959 & 2011 Nov 6  & (3,3)\\
08311--2459 & OH119 \& OH79 & 1342230967 & 2011 Oct 14  & (3,3)\\
10378+1109  & OH119 \& OH79 & 1342232315 & 2011 Nov 14 & (3,3)\\
11095--0238 & OH119 \& OH79 & 1342232612 & 2011 Nov 22 & (3,3)\\
12071--0444 & OH119 \& OH79 & 1342234994 & 2011 Dec 20 & (3,3)\\
3C273       & OH119 \& OH79 & 1342235707 & 2011 Dec 29 & (3,3)\\ 
13451+1232  & OH119 \& OH79 & 1342236886 & 2012 Jan 9   & (3,3)\\
Mrk463      & OH119 \& OH79 & 1342236982 & 2012 Jan 8   & (3,3)\\
15462--0450 & OH119 \& OH79 & 1342238135 & 2012 Jan 22  & (3,3)\\
16090--0139 & OH119 \& OH79 & 1342238907 & 2012 Feb 10 & (3,3)\\
19254--7245 & OH119 \& OH79 & 1342231722 & 2011 Oct 31  & (2,2)+(2,3)+(3,3)\\
20087--0308 & OH119 \& OH79 & 1342232302 & 2011 Nov 13 & (3,3)\\
20100--4156 & OH119 \& OH79 & 1342216371 & 2011 Mar 18    & (3,3)\\
20414--1651 & OH119 \& OH79 & 1342217908 & 2011 Apr 3     & (3,3)\\
23230--6926 & OH119 \& OH79 & 1342234997 & 2011 Dec 21 & (3,3)\\
23253--5415 & OH119 \& OH79 & 1342235703 & 2011 Dec 28 & (3,3)
\enddata

\tablenotetext{a}{Spaxel (3,3) is the central spaxel in the $5\times5$
footprint.} 
\tablenotetext{b}{Observed as part of program {\tt OT1\_dweedman\_1}. The HERUS
  observation is mispointed.}
\tablenotetext{c}{Observed as part of program {\tt OT1\_sveilleu\_3}. The HERUS
  observation is mispointed.}

\end{deluxetable*}

A promising avenue for diagnosing the properties of AGN-driven outflows, 
with the eventual aim of understanding the feedback phenomenon, is to 
examine outflows of {\itshape molecular} gas. In ULIRGs such outflows 
may be expected to trace the earliest stages of AGN feedback activity, 
while the AGN itself is still obscured. They also provide the only 
direct evidence that the fuel for star formation is being removed.
Moreover, molecular outflows have
been observed previously in IR-luminous galaxy mergers 
\citep{sturm11,chung11,fischer10,sakamoto09,walter02,baan89}. 
The observational study of such 
outflows is however still in its infancy; we currently lack even a
broad phenomenological picture of how and when outflow activity is 
seen in molecular lines, and what other properties of the merger the 
molecular line profiles show relationships with. 

The Herschel ULIRG Survey (HERUS) is well situated to examine the 
range of properties of an important molecular tracer, hydroxyl (OH), 
in local IR-luminous mergers. HERUS has obtained spectroscopic 
observations of two ground-state doublets of OH, at 79\,$\mu$m and 
119\,$\mu$m, in 24 local ULIRGs. 
In this paper we present a summary of the full OH data set, analyze 
the velocity profiles, and compare them to line profiles of mid and 
far-IR fine-structure lines. Modeling of the sources will be presented 
in subsequent HERUS papers. In Sect.\,2 we review observations of
OH in the literature. In Sect.\,3 we present new observations we
obtained using Herschel. In Sect.\,4 we analyze the velocity profiles
of the OH profiles and compare them to line profiles of mid and 
far-IR fine-structure lines. In Sect.\,5 we discuss and summarize 
our results.
Throughout this paper we assume H$_0$ = 67.3 km s$^{-1}$ Mpc$^{-1}$, 
$\Omega_M$ = 0.315, and $\Omega_\Lambda$ = 0.685 \citep{planck13}.

\section{Background:  observations and properties of OH}

The first far infrared detections of OH in the interstellar medium were 
made by \cite{storey81} using the Kuiper Airborne Observatory: they 
measured the two 119\,$\mu$m $\Lambda$-doubled lines between the
ground and first excited states, which they discovered in absorption 
from Sgr\,B2 and in emission from the shock in Orion\,KL. This line 
is the strongest OH line, being from a ground-state absorption, but 
there are altogether fourteen far-infrared lines between 34\,$\mu$m 
and 163\,$\mu$m, arising among the lowest eight rotational levels 
of OH. The transitions, which arise from levels lying 120 to 618\,K
above ground, are strong. The dipole moment for OH is large, 1.668 
Debye (for comparison the CO dipole moment is 0.112 Debye), so that
the radiative rates for OH transitions are generally fast (for example, 
the Einstein A coefficient for the 119\,$\mu$m doublet lines is about 
0.136\,s$^{-1}$, as compared to 7.2$\times$10$^{-8}$\,s$^{-1}$ for CO J=1-0). 
Critical densities for OH are high (of the order 10$^9$\,cm$^{-3}$), 
making these transitions sensitive to radiative pumping. The FIR OH 
transitions also include cross-ladder lines whose radiative rates are 
one hundred times weaker, providing a dataset of neighboring, far IR 
lines which frequently include both optically thin and very optically 
thick features. The OH analyses have an additional resource from which 
to draw: the strong hyper-fine radio wavelength transitions OH has in 
its ground-state (1665 and 1667 MHz; 18cm), which have been extensively 
observed \citep[e.g.][]{baan85,baan89,henkel90}.

ISO detected the remaining OH infrared lines from Galactic and/or 
extragalactic sources. \cite{sylvester97} used them to show
conclusively that the 18\,cm radio maser emission in evolved stars 
is pumped by absorption of the 34\,$\mu$m dust continuum, while 
\cite{skinner97} proved the effectiveness of the IR pumping of mega-masers 
in ULIRGs in the case of Arp220.  The ISO analyses and modeling of the 
OH line soon discovered their power and complexity, especially in 
conjunction with modeling the FIR fine-structure lines. 
\citet[][]{fischer99}, \citet[][NGC253]{bradford99}, \citet[][IRAS
20100--4156 and 3Zw35]{kegel99}, \citet[][Arp220]{gonzalez04}, \citet[][]{smith04},
\citet[][NGC1068]{spinoglio05}, and \citet[][Mrk231]{gonzalez08}
found the OH lines either mostly in absorption (Arp220; IRAS\,20100--4156), 
mostly in emission (NGC1068), or with a mix (Mrk 231; NGC 253).  
The models, which had to 
fit the dust continua and atomic fine structure lines, usually required 
a small, warm nuclear region, a cooler extended region, and sometimes a 
third component as well, with evidence for collisional excitation, 
radiative pumping from the presence of warm dust continuum photons, and 
also signs of XDR excitation (later suggested for pumping the CO 
infrared lines as well).

Herschel observations have dramatically improved the dataset, not only 
in terms of sensitivity, but also because it resolved the line shapes 
in many cases. \cite{fischer10} reported the remarkable discovery that 
some of the OH lines show P-Cygni profiles, suggesting massive outflows, 
and \cite{sturm11} and \cite{gonzalez10,gonzalez12,gonzalez13} 
among others have begun to exploit and model the implications of these 
data.  Outflows are expected to play a pivotal role in the evolutionary 
history of galaxies, and indeed radio observations of CO and HCN
\citep{aalto12,feruglio10} are now making inroads into this topic.  It 
appears that the OH molecule plays a remarkable diagnostic role in that 
the combination of operative excitation mechanisms and optical depths 
for its many infrared lines allow it to be a much broader sensor of 
the various conditions in galaxies (nuclear regions, star formation 
complexes, outflows, etc.) than, for example, is CO, yet OH can have an 
abundance not much smaller.

\section{Observations}

The Herschel ULIRG Survey (HERUS; Farrah et al. in prep.) is a 
far-IR photometric and spectroscopic atlas of the low-redshift 
(z$<$0.262) ULIRG population with
Herschel, and at 250 hours, was the largest extragalactic program 
in the Herschel Open Time 1 call. The sample comprises all 40 
ULIRGs from the IRAS-PSCz with IRAS 60\,$\mu$m flux densities 
$\geq$ 2 Jy, plus IRAS\,00397--1312 (1.8\,Jy), IRAS\,07598+6508 
(1.7\,Jy) and IRAS\,13451+1232 (1.9\,Jy). All objects have been 
observed with the Infrared Spectrograph (IRS) onboard Spitzer. 
HERUS obtained SPIRE-FTS spectra for 29/43 sources and SPIRE
photometry for 32/43 sources.
PACS spectroscopy was obtained for any target not part of the 19 ULIRGs 
included in the SHINING program. The HERUS PACS subsample hence comprises
24 sources, whose observation details are shown in Table\,\ref{tbl-0}.
Among these sources is the blazar 3C273, a ULIRG by luminosity, but 
otherwise not a late-stage merger. The source will hence not be
considered in Sect.\,\ref{vmax_balnicity_section} and beyond, where 
we discuss evolutionary aspects of dusty mergers.

\begin{deluxetable*}{lrrrrrrrr}
\tablewidth{0pt}
\tablecaption{Basic observational quantities for the HERUS sample\label{tbl-1}}
\tablehead{\colhead{Galaxy} & \colhead{Redshift\tablenotemark{a}} & \colhead{log $L_{\rm IR}$\tablenotemark{b}/$L_{\odot}$} &  \colhead{$f_{30}$/$f_{15}$\tablenotemark{c}} & \colhead{$\alpha$\tablenotemark{d}} & \colhead{log $L_{{\rm AGN},bol}$\tablenotemark{e}/$L_{\odot}$} & \colhead{SFR\tablenotemark{f}} & \colhead{S$_{sil}$\tablenotemark{g}} & \colhead{EQW62\tablenotemark{h}}\\
\colhead{} & \colhead{} & \colhead{} & \colhead{} & \colhead{} & \colhead{} & \colhead{(M$_{\odot}$ yr$^{-1}$)} & \colhead{} & \colhead{($\mu$m)}}
\startdata
00188-0856 &  0.1287 &   12.39 &    9.05 &  0.51 &   12.16$^{+ 0.11}_{-0.15}$ &   119$^{+  36}_{ -36}$ & -2.6 &  0.066 \\
00397-1312 &  0.2617 &   12.90 &    6.29 &  0.67 &   12.79$^{+ 0.09}_{-0.11}$ &   263$^{+ 119}_{-119}$ & -2.9 &  0.026 \\
01003-2238 &  0.1179 &   12.32 &    3.88 &  0.83 &   12.30$^{+ 0.07}_{-0.09}$ &    36$^{+  31}_{ -31}$ & -0.8 &  0.040 \\
Mrk1014 &  0.1631 &   12.62 &    5.65 &  0.71 &   12.53$^{+ 0.08}_{-0.10}$ &   121$^{+  62}_{ -62}$ &  0.2 &  0.080 \\
03158+4227 &  0.1346 &   12.63 &    9.93 &  0.47 &   12.36$^{+ 0.12}_{-0.17}$ &   227$^{+  63}_{ -63}$ & -3.1 &  0.059 \\
03521+0028 &  0.1519 &   12.52 &   20.21 &  0.06 &   11.39$^{+ 0.52}_{-2.00}$ &   309$^{+  21}_{ -49}$ & -1.3 &  0.356 \\
06035-7102 &  0.0795 &   12.22 &    7.50 &  0.60 &   12.06$^{+ 0.10}_{-0.13}$ &    66$^{+  24}_{ -24}$ & -1.5 &  0.087 \\
06206-6315 &  0.0921 &   12.23 &   10.72 &  0.43 &   11.92$^{+ 0.13}_{-0.19}$ &    96$^{+  25}_{ -25}$ & -1.6 &  0.183 \\
07598+6508 &  0.1487 &   12.50 &    2.74 &  0.91 &   12.52$^{+ 0.04}_{-0.08}$ &    28$^{+  47}_{ -28}$ &  0.1 &  0.006 \\
08311-2459 &  0.1005 &   12.50 &    4.36 &  0.79 &   12.46$^{+ 0.08}_{-0.09}$ &    65$^{+  47}_{ -47}$ & -0.5 &  0.139 \\
10378+1109 &  0.1363 &   12.31 &   13.55 &  0.30 &   11.85$^{+ 0.17}_{-0.30}$ &   142$^{+  30}_{ -30}$ & -2.1 &  0.091 \\
11095-0238 &  0.1063 &   12.28 &    9.56 &  0.49 &   12.03$^{+ 0.12}_{-0.16}$ &    97$^{+  28}_{ -28}$ & -3.3 &  0.037 \\
12071-0444 &  0.1289 &   12.41 &    4.95 &  0.75 &   12.35$^{+ 0.08}_{-0.10}$ &    63$^{+  38}_{ -38}$ & -1.4 &  0.087 \\
3C273 &  0.1586 &   12.80 &    1.38 &  1.00 &   12.86$^{+ 0.00}_{-0.07}$ & $<$94 &  0.1 & $<$0.007 \\
Mrk231\tablenotemark{i} &  0.0422 &   12.55 &    4.10 &  0.81 &   12.52$^{+ 0.07}_{-0.09}$ &    67$^{+  53}_{ -53}$ & -0.7 &  0.010 \\
13451+1232 &  0.1218 &   12.32 &    4.01 &  0.82 &   12.29$^{+ 0.07}_{-0.09}$ &    38$^{+  31}_{ -31}$ & -0.3 & $<$0.022 \\
Mrk463 &  0.0508 &   11.79 &    1.89 &  0.98 &   11.84$^{+ 0.01}_{-0.07}$ &     1$^{+   9}_{  -1}$ & -0.4 & $<$0.005 \\
15462-0450 &  0.1003 &   12.24 &    6.94 &  0.63 &   12.10$^{+ 0.09}_{-0.12}$ &    64$^{+  26}_{ -26}$ & -0.4 &  0.062 \\
16090-0139 &  0.1336 &   12.55 &   10.63 &  0.43 &   12.25$^{+ 0.13}_{-0.18}$ &   200$^{+  53}_{ -53}$ & -2.5 &  0.071 \\
19254-7245 &  0.0619 &   12.09 &    5.17 &  0.74 &   12.02$^{+ 0.08}_{-0.10}$ &    32$^{+  18}_{ -18}$ & -1.3 &  0.066 \\
20087-0308 &  0.1057 &   12.42 &   16.07 &  0.20 &   11.79$^{+ 0.24}_{-0.57}$ &   209$^{+  39}_{ -39}$ & -1.8 &  0.367 \\
20100-4156 &  0.1296 &   12.67 &   14.45 &  0.27 &   12.16$^{+ 0.19}_{-0.36}$ &   343$^{+  70}_{ -70}$ & -2.7 &  0.089 \\
20414-1651 &  0.0869 &   12.22 &   22.86 &  0.00\tablenotemark{j} & $<$11.46 &   165$^{+   0}_{ -24}$ & -1.6 &  0.570 \\
23230-6926 &  0.1064 &   12.37 &   13.25 &  0.32 &   11.93$^{+ 0.17}_{-0.28}$ &   160$^{+  35}_{ -35}$ & -2.1 &  0.324 \\
23253-5415 &  0.1298 &   12.36 &   15.36 &  0.23 &   11.78$^{+ 0.22}_{-0.46}$ &   176$^{+  34}_{ -34}$ & -1.5 &  0.235 \\
\enddata
\tablenotetext{a}{Determined from far-IR fine-structure lines.}
\tablenotetext{b}{L$_{\rm IR}$=$L$(8--1000\,$\mu$m) integrated from SEDs consisting of Spitzer-IRS spectra, and IRAS, ISO-PHT and Herschel-SPIRE photometry.}
\tablenotetext{c}{Ratio of 30\,$\mu$m to 15\,$\mu$m continuum.}
\tablenotetext{d}{Fraction of L$_{\rm bol}$ contributed by the AGN, $\alpha$, derived using method \#6 of \cite{veilleux09}. We assume the uncertainty in $\alpha$ to be $\pm$0.15.}
\tablenotetext{e}{Derived using method \#6 by \cite{veilleux09}. This method uses $f_{30}$/$f_{15}$ to infer $\alpha$.}
\tablenotetext{f}{Using SFR=(1-$\alpha$)$\times$10$^{-10}$\,L$_{\rm IR}$ as adopted by \cite{sturm11}.}
\tablenotetext{g}{Silicate strength \citep{spoon07}: $>$0 for silicate emission features, $<$0 for silicate absorption features.}
\tablenotetext{h}{Ice-corrected equivalent width of the 6.2\,$\mu$m PAH feature \citep{spoon07}.}
\tablenotetext{i}{Observed as part of the SHINING sample \citep{fischer10}}
\tablenotetext{j}{Methods \#3--6 of \cite{veilleux09} estimate $\alpha$=0. Only method \#2, based on the detection of 14.32\,$\mu$m [NeV] \citep{farrah07}, estimates $\alpha$=0.60.}
\end{deluxetable*}

The PACS spectroscopy observations for the HERUS program were 
performed between March 18, 2011 and April 8, 2012 (OD673--OD1060). 
The PACS integral-field spectrometer samples the spatial direction 
with 25 pixels and the spectral direction with 16 pixels. Each 
spectral pixel ``sees'' a distinct wavelength range that is scanned 
by varying the grating angle, and the combination of the 16 ranges 
makes the total spectrum. The $5\times5$ spatial pixels (``spaxels'') 
constitute a field-of-view of 47$"\times$47$"$, much larger than 
the far-IR source sizes. The coordinates were chosen to center the 
object in the central spaxel. According to the PACS Observer's 
Manual\footnote{{\it http://herschel.esac.esa.int/Docs/PACS/html/pacs\_om.html}}, 
the point spread function full width at half maximum (FWHM) is 
$\approx9.5"$ between 55 and 100\,$\mu$m, and it increases to about 
$\approx14"$ at 200\,$\mu$m (the longest observed wavelength in
our sample is 199\,$\mu$m).

Observations were done in range spectroscopy mode with a spectral 
range somewhat larger than that used for line spectroscopy mode. 
For a given feature, the width of the spectral range was chosen to 
accommodate possible uncertainties in the redshift, as well as accounting 
for blue wings and blue shifts and for intrinsic line broadening. 
The chop/nod 
observation mode was used, in which the source is observed by 
alternating between the on-source position and a clean off-source 
position. The smallest throw ($\pm1.5\arcmin$) was used to reduce 
the effect of field-rotation between the two chop positions. Two 
nod positions are used in order to completely eliminate the 
telescope background emission.

We observed all the HERUS objects in the following lines: 
52\,$\mu$m [O III], 57\,$\mu$m [N III], 63\,$\mu$m [O I], 
145\,$\mu$m [O I], and 158\,$\mu$m [C II]. In addition, the 
122\,$\mu$m [N II] line was observed together with the 119\,$\mu$m
OH doublet (OH119). Since the PACS spectrometer observes 
simultaneously in 2 bands, the 79\,$\mu$m OH doublet (OH79) 
was observed in the blue band while 158\,$\mu$m [C II] was 
observed in the red band. Results
for our fine-structure line observations will be presented by
\cite{farrah13}.

The data reduction was performed in HIPE {\tt 8.0} \citep{ott10}
using the default chop/nod pipeline script. The level 1 product 
(calibrated in flux and in wavelength, with bad pixels masks according 
to the HIPE reduction criteria) was then exported and processed by 
our in-house PACSman tool \citep{lebouteiller12} for further data 
flagging, for spectral rebinning and empirical error estimates. 

In this paper we focus on the spectra of the brightest spaxel in the
field -- typically spaxel (3,3) -- and perform a point source flux 
correction. There are, however, exceptions. For IRAS\,19254--7245 
(``the SuperAntennae'') we intended to point to its southern
nucleus. An error on the input coordinates meant that this nucleus
fell between spaxels (2,2), (2,3) and (3,3). Since all three show equal 
continuum brightness we coadded their spectra. See Table\,\ref{tbl-0}.

Also for two other sources in our sample inaccurate coordinates were 
used to obtain our spectra. IRAS\,03158+4227 was mispointed by about 
half a spaxel, and IRAS\,07598+6508 by about two spaxels, thereby 
placing the source at an edge spaxel. For the latter source we 
therefore substituted the HERUS observations of OH79 and OH119 by 
those obtained by \cite{sargsyan12} (program {\tt OT1\_dweedman\_1})
and Veilleux et al. (program {\tt OT1\_sveilleu\_3}), respectively.

The redshifts for our sources were 
obtained from observations of 158\,$\mu$m [C II], 122\,$\mu$m 
[N II] and 63\,$\mu$m [O I], and have been tabulated in 
Table\,\ref{tbl-1}.

\section{Analysis}

Our spectra of the 119\,$\mu$m OH (OH119) and 120\,$\mu$m $^{18}$OH 
($^{18}$OH120) doublets are shown in Fig.\,\ref{oh119_spectra}. Each 
spectrum covers a velocity range of -2,500 to +5,000 km/s with respect 
to the bluest of the two OH119 doublet components
($\lambda_{rest}$=119.233\,$\mu$m; $\Delta$$v_{\rm doublet}$=520\,km s$^{-1}$). 
The redshifts used to define systemic velocity ($v_{\rm sys}$) for this 
line are given in Table\,\ref{tbl-1}. Apart from OH119 and $^{18}$OH120 
several other species have lines in this range: CH$^+$ (at 
$v$=+1546\,km s$^{-1}$) and CH at --1330 and --2080\,km s$^{-1}$. 
The latter species is not detected in any of our sources. Here we 
will assume clean continuum to exist at $v$$<$--2,000\,km s$^{-1}$ 
and at $v$$>$+2,500\,km s$^{-1}$. The local continuum resulting from
second order polynomial interpolation between these two continuum
ranges is shown as a dashed red line in Fig.\,\ref{oh119_spectra}.

The corresponding spectra of the 79\,$\mu$m OH (OH79) and $^{18}$OH 
($^{18}$OH79) doublets are shown in Fig.\,\ref{oh79_spectra}. The spectra 
cover a velocity range of -2,500 to +2,000\,km s$^{-1}$ with respect to 
the bluest of the two OH79 doublet components
($\lambda_{rest}$=79.118\,$\mu$m; $\Delta$$v_{\rm doublet}$=240\,km s$^{-1}$),
and have been rebinned to channels of $\sim$80\,km s$^{-1}$ velocity
width to make up for poor S/N. Unlike the OH119 and $^{18}$OH120 doublets, 
which are separated by $\sim$1,900\,km s$^{-1}$, 
the two 79\,$\mu$m OH isotope doublets are strongly blended at 
the resolution of the instrument. Apart from these OH doublets
the spectral range also includes the 78.742\,$\mu$m H$_2$O 
4$_{23}$--3$_{12}$ transition (at $v$=--1425\,km s$^{-1}$), which 
may be detected in several of our sources (most clearly in 
IRAS\,20087--0308, 20100--4156 and 23230--6923). Given the poor
S/N of the OH79 spectra, the continuum was fitted using first order
polynomials only.

\begin{deluxetable}{ll}
\tablewidth{0pt}
\tablecaption{OH lines log\label{tbl-4}}
\tablehead{
\colhead{Line} & \colhead{$\lambda$\tablenotemark{a}}\\
\colhead{} & \colhead{($\mu$m)}}
\startdata
$^{18}$OH $\Pi_{1/2}$ -- $\Pi_{3/2}$ $\frac{1^-}{2}$ -- $\frac{3^+}{2}$ & 79.083 \\
$^{16}$OH $\Pi_{1/2}$ -- $\Pi_{3/2}$ $\frac{1^-}{2}$ -- $\frac{3^+}{2}$ & 79.118 \\
$^{18}$OH $\Pi_{1/2}$ -- $\Pi_{3/2}$ $\frac{1^+}{2}$ -- $\frac{3^-}{2}$ & 79.147 \\
$^{16}$OH $\Pi_{1/2}$ -- $\Pi_{3/2}$ $\frac{1^+}{2}$ -- $\frac{3^-}{2}$ & 79.181 \\
$^{16}$OH $\Pi_{3/2}$ -- $\Pi_{3/2}$ $\frac{5^-}{2}$ -- $\frac{3^+}{2}$ & 119.233 \\
$^{16}$OH $\Pi_{3/2}$ -- $\Pi_{3/2}$ $\frac{5^+}{2}$ -- $\frac{3^-}{2}$ & 119.441 \\
$^{18}$OH $\Pi_{3/2}$ -- $\Pi_{3/2}$ $\frac{5^-}{2}$ -- $\frac{3^+}{2}$ & 119.964 \\
$^{18}$OH $\Pi_{3/2}$ -- $\Pi_{3/2}$ $\frac{5^+}{2}$ -- $\frac{3^-}{2}$ & 120.171 
\enddata
\tablenotetext{a}{The tabulated rest wavelengths are averages of the rest wavelengths of
the two blended hyper-fine lines.}
\end{deluxetable}

\subsection{Examination of the OH119 profiles}

The spectra in Fig.\,\ref{oh119_spectra} show a great variety
in OH119 profiles, ranging from pure absorption profiles 
(e.g. IRAS\,06206--6315 and 20087--0308), via P-Cygni profiles
(e.g. IRAS\,06035--7102 and 12071--0444), to pure emission 
profiles (e.g. IRAS\,00397--1312 and 13451+1232). The normalized
versions of these spectra, shown in
Fig.\,\ref{normalized_oh119_spectra}, further reveal the
contrast and equivalent width of the OH119 features to vary
greatly. The deepest absorption feature is seen in IRAS\,03158+4227,
and the strongest emission feature in IRAS\,13451+1232. 
The relative strength of the $^{18}$OH120 to the OH119 doublets also 
appears to vary within the sample. $^{18}$OH120 seems absent in 
IRAS\,08311--2459 and very strong in IRAS\,20100--4156.

The double peaked nature of the OH119 doublet is evident 
in most spectra, although there are some sources, like 
IRAS\,10378+1109 and 13451+1232, where it is not self-evident. 
Perhaps in these sources this signature
is washed out due to a strong velocity gradient, or due to
the existence of several velocity components. In our sample 
the velocity range over which the OH absorption and/or
emission occurs varies greatly. The narrowest velocity 
range is found in IRAS\,20414--1651, where both OH119
doublet lines appear to be unresolved. The widest velocity 
field likely occurs in IRAS\,03158+4227. Note that in the 
latter source the lowest blue shifted outflow velocity 
amounts to $\sim$400\,km s$^{-1}$, whereas in IRAS\,20100--4156, 
for example, also quiescent gas (gas at v$\approx$0) appears to be present.
In Mrk1014 the entire OH119 profile, seen in emission, seems 
to be red shifted by 200--300\,km s$^{-1}$. Some of our sources 
further show a lack of red shifted emitting high-velocity gas 
at similar velocities as the blue shifted absorbing gas. 
The clearest example may be the OH119 profile of IRAS\,01003--2238, 
since this OH119 profile is not contaminated 
by $^{18}$OH120 emission or absorption. The significance 
of this disparity will be discussed in Sect.\,\ref{oh119quali_section}.

Due to the relatively small separation in velocity space between
the four OH and $^{18}$OH lines ($\sim$500\,km s$^{-1}$ between
doublet lines and $\sim$2000\,km s$^{-1}$ between $^{16}$OH and 
$^{18}$OH isotopes), it is generally not straightforward to 
extract basic kinematic information from the blended profiles 
without involving model fits. If, for instance, OH and $^{18}$OH 
exist out to velocities more blueshifted than -1000\,km s$^{-1}$ 
and more redshifted than +1000\,km s$^{-1}$, 
the $^{18}$OH120 complex will bleed into the OH119 complex. 
Hence, only within 500\,km s$^{-1}$ of the highest blue shifted 
velocity of the 119.233\,$\mu$m OH line and within 500\,km s$^{-1}$
of the highest red shifted velocity of the 120.171\,$\mu$m $^{18}$OH 
line one can be certain that the other doublet line and the 
other isotope do not contribute to the spectral structure. 
We will use this aspect in Sect.\,\ref{vmax_section} to 
determine the highest velocity blue shifted OH gas.

\begin{figure*}
\epsscale{.99}
\includegraphics[scale=1.0]{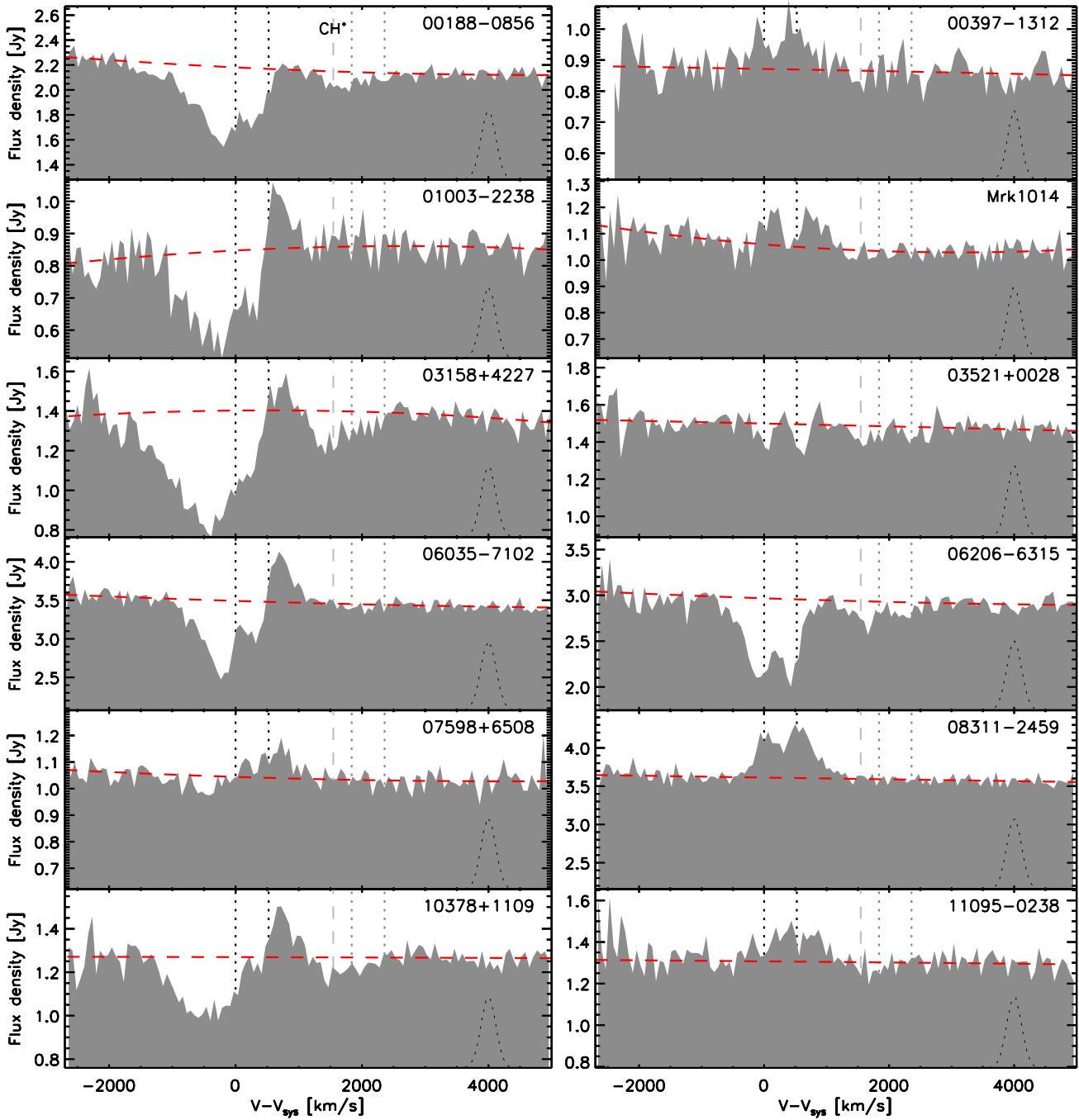}
\caption{Herschel-PACS spectra of the partially blended 119\,$\mu$m 
$^{16}$OH and $^{18}$OH doublets for ULIRGs observed in the HERUS 
survey. All spectra have the same flux dynamic range. Systemic 
velocity is defined as the rest wavelength of the $^{16}$OH 
$\Pi_{3/2}$ -- $\Pi_{3/2}$ $\frac{5^-}{2}$ -- $\frac{3^+}{2}$ 
transition at 119.233\,$\mu$m. 
The rest velocities of the $^{16}$OH and $^{18}$OH lines (see 
Table\,\ref{tbl-4}) are indicated by black and grey dotted 
vertical lines, respectively. Potential contamination by
CH$^+$ (119.848\,$\mu$m) is indicated by a light grey dashed 
vertical line at $v$=+1543\,km s$^{-1}$.
The adopted local continuum is shown in red. The spectral 
resolution is shown by a dashed black gaussian profile in 
the right bottom corner of each panel. 
\label{oh119_spectra}}
\end{figure*}

\begin{figure*}
\addtocounter{figure}{-1}
\epsscale{.99}
\includegraphics[scale=1.0]{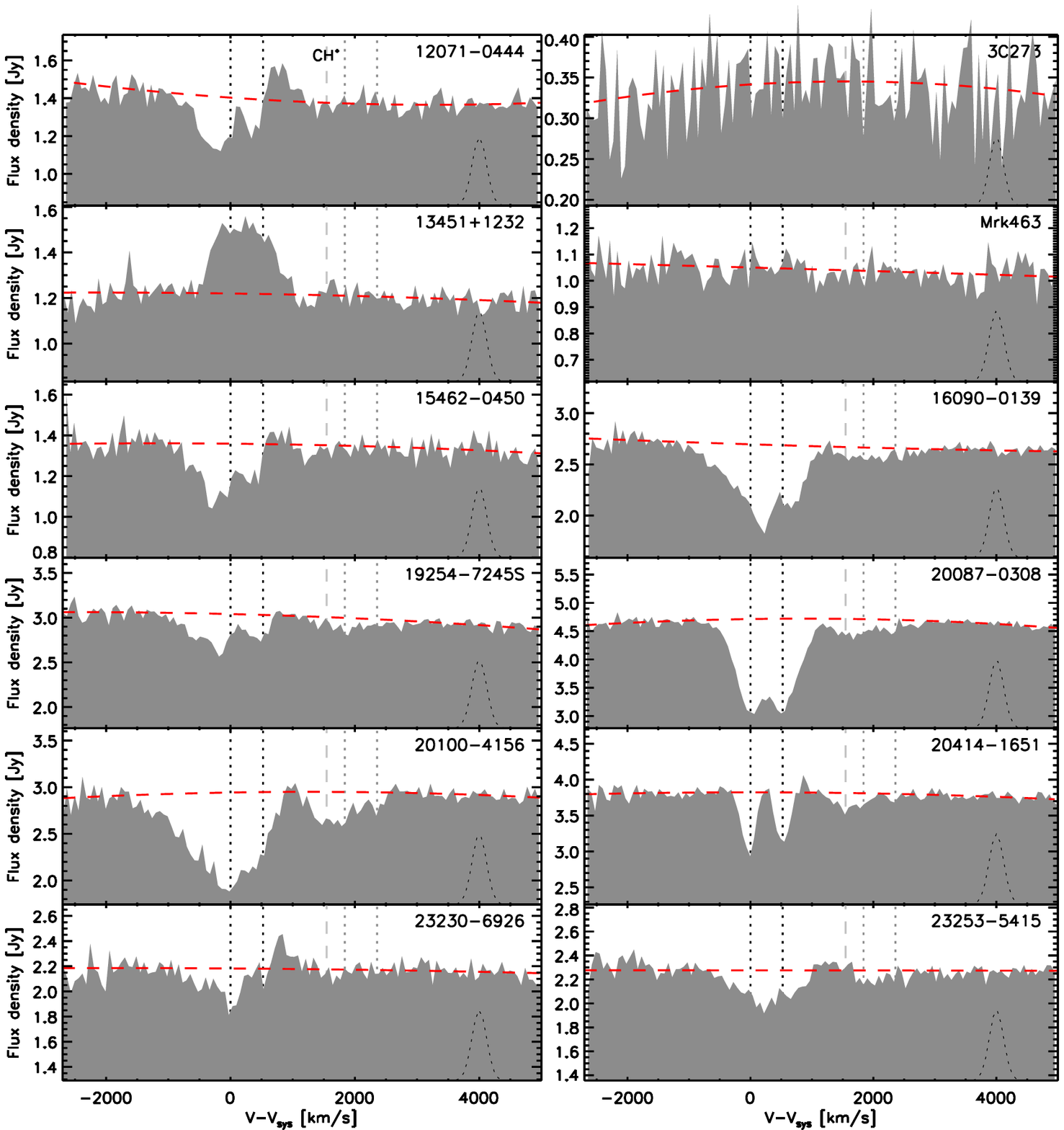}
\caption{119\,$\mu$m OH (continued).}
\end{figure*}

\begin{figure*}
\includegraphics[scale=1.0]{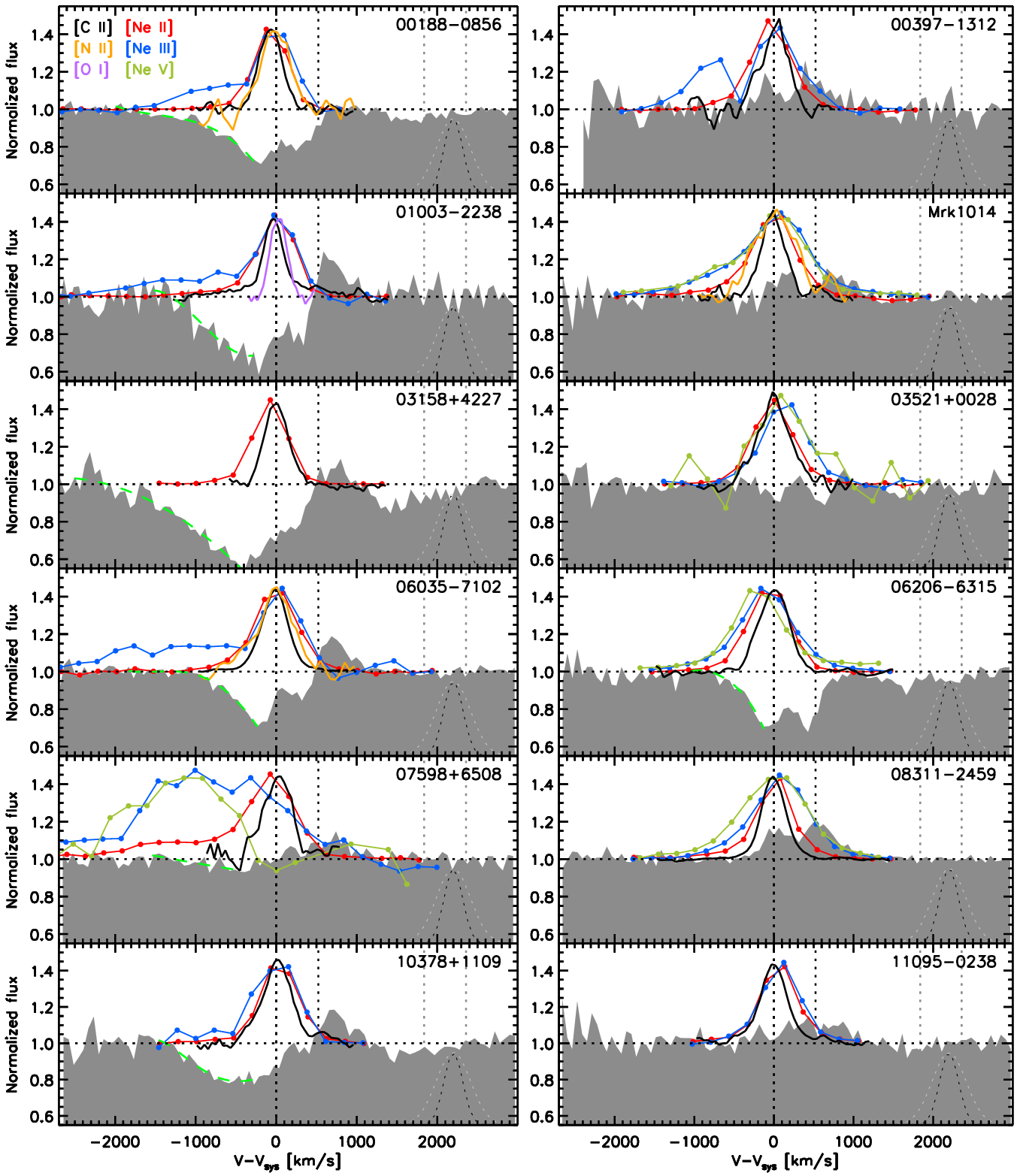}
\caption{Comparison of the normalized 119\,$\mu$m OH doublet 
profile to the scaled, smoothed emission line profiles of mid 
and far-IR fine-structure lines. The rest velocities of the 
$^{16}$OH and $^{18}$OH lines (see Table\,\ref{tbl-4}) are 
indicated by black and grey dotted lines, respectively.
Line profiles of 12.81\,$\mu$m [NeII], 15.56\,$\mu$m [NeIII]
and 14.32\,$\mu$m [NeV] are color-coded as red, blue and
green dotted lines, respectively. [C II] is shown in black,
122\,$\mu$m [N II] in orange, and 63\,$\mu$m [O I] in purple.
The B-spline fit to the blue wing of the OH119 profile is
shown as a green dashed curve. The spectral resolution for 
the OH119 line is shown by a dashed black gaussian 
profile in the right bottom corner of each panel. The
Spitzer-IRS-SH spectral resolution is shown as a light 
grey profile.
\label{normalized_oh119_spectra}}
\end{figure*}

\begin{figure*}
\addtocounter{figure}{-1}
\includegraphics[scale=1.0]{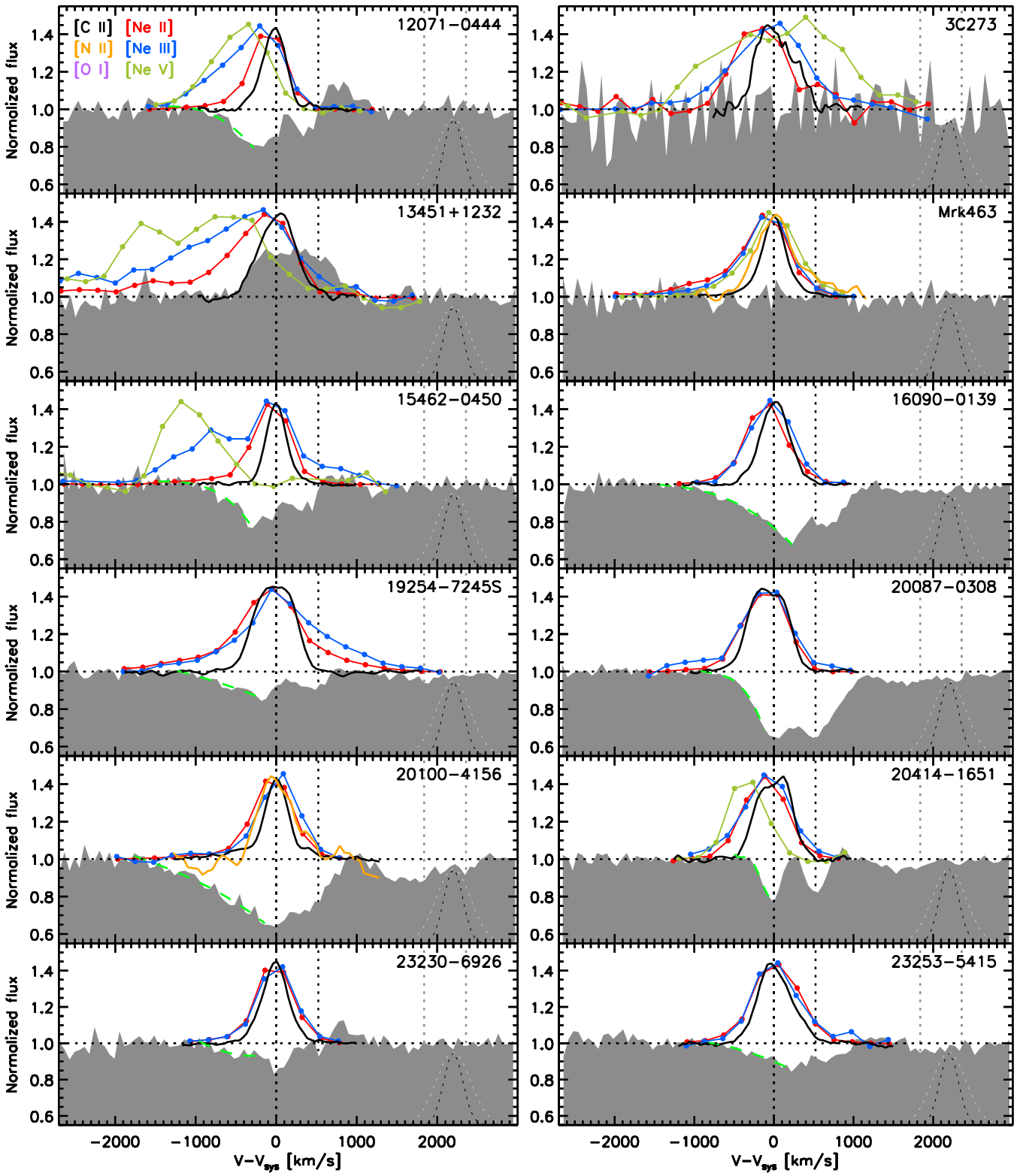}
\caption{Comparison of normalized and scaled line profiles.
(continued).}
\end{figure*}

\subsection{Examination of the OH79 profiles}

Our sample of OH79 spectra (shown in Fig.\,\ref{oh79_spectra})
generally has lower S/N than the corresponding sample of OH119 spectra, 
even after rebinning to channels of 80\,km s$^{-1}$. Combined 
with a lower feature contrast \citep[typically $<$10\% of the 
continuum;][compared to up to 45\% for OH119]{sturm11} this 
makes examination of individual OH79 profiles more challenging 
than that of OH119 profiles.

Our sample of OH79 profiles lacks the deep blue shifted OH 
absorption components that are present in many of our OH119 
profiles. Instead, the most clearly detected feature in the 
OH79 spectra (in about 50\% of the sample) is a red shifted 
OH emission component. These emission components range from 
pure emission profiles (IRAS\,08311--2459, 11095--0238, and 
13451+1232), via red shifted P Cygni components (IRAS\,01003--2238, 
03158+4227, 06035--7102, 10378+1109, 12071--0444, 15462--0450, 
20100--4156), to pure absorption profiles (IRAS\,20087--0308). 
The majority of our OH79 profiles are hard to classify.

The differences between the overall appearance of the 
OH119 and OH79 profile samples may seem surprising, given 
that both doublet complexes are the result of transitions 
between the $\Pi_{3/2}$ ground-state level and the lowest 
excited level in the $\Pi_{1/2}$ and $\Pi_{3/2}$ ladders. 
There are, however, important differences.
The Einstein B coefficients for the OH79 transitions are 
about 40 times lower than for OH119 \citep{fischer10}, implying 
higher optical depths in the OH119 than in the OH79 doublet.
The latter doublet is therefore far better suited to measure 
the OH column density than the optically thick OH119 doublet,
but it also requires far higher S/N to detect it.

There are, however, conditions under which it is not straightforward 
to measure the OH column density from the OH79 doublet, and this 
depends on which of the pathways available to radiatively
pump the molecule dominates. Unlike the OH119 doublet, 
which is an in-ladder ground-state transition in the $\Pi_{3/2}$ 
ladder, the OH79 doublet is a cross-ladder transition 
between the $\Pi_{3/2}$ and $\Pi_{1/2}$ ladders.
This distinction is important as there are two other cross-ladder 
ground-state transitions which can very effectively pump the 
$\Pi_{1/2}$ levels above the OH79 upper level by absorbing 
35\,$\mu$m or 53\,$\mu$m IR dust continuum photons 
\citep[see e.g.][]{spinoglio05,gonzalez12}. A similar 
ground-state pumping mechanism for levels above the OH119 
upper level does not exist in the $\Pi_{3/2}$ ladder.
As there are no strong\footnote{The Einstein A coefficient
for a cascade from the $\Pi_{1/2}$ J=3/2 level to the 
$\Pi_{1/2}$ J=1/2 level (163\,$\mu$m OH doublet) is seven 
times larger than for a cross-ladder cascade to the OH119 
upper level in the $\Pi_{3/2}$ ladder (96\,$\mu$m OH doublet).} 
(or efficient) cross-ladder transitions 
from the $\Pi_{1/2}$ ladder to the $\Pi_{3/2}$ OH119 upper level, 
most radiative cascades from the $\Pi_{1/2}$ ladder to the ground-state 
(in the $\Pi_{3/2}$ ladder) take place through the OH79 transition.
So, while the OH119 upper level is almost exclusively 
excited through the absorption of a 119\,$\mu$m 
photon\footnote{Collisional excitation in
a dense/warm environment is of course also a viable way 
to populate the OH119 upper level.}, the OH79 upper level 
may have become populated not only by absorption of 79\,$\mu$m 
but also 35 and 53\,$\mu$m IR dust continuum photons.
And this opens the possibility for OH79 to be purely 
in emission while OH119 shows an absorption component.
In our sample we cannot identify a single case in which the
nature of the OH79 profile differs from that of the OH119 
profile. I.e., for as far as the S/N of our spectra permits, 
none of the sources for which OH119 shows an absorption
component show OH79 purely in emission, or vice versa. 

Further direct comparisons between the OH79 and OH119
profiles are complicated by the different doublet spacings
and the different isotope spacings for these OH complexes.
For instance, the red emission component of an OH79 P Cygni 
profile always appears closer to systemic velocity (by 300
to 500\,km s$^{-1}$; e.g. IRAS\,01003--2238, 06035--7102,
and 10378+1109) than the red emission component of the 
corresponding OH119 doublet. This substantial difference
can be partially explained by the differences in doublet 
spacings between OH79 and OH119.

\begin{figure*}
\includegraphics[scale=1.0]{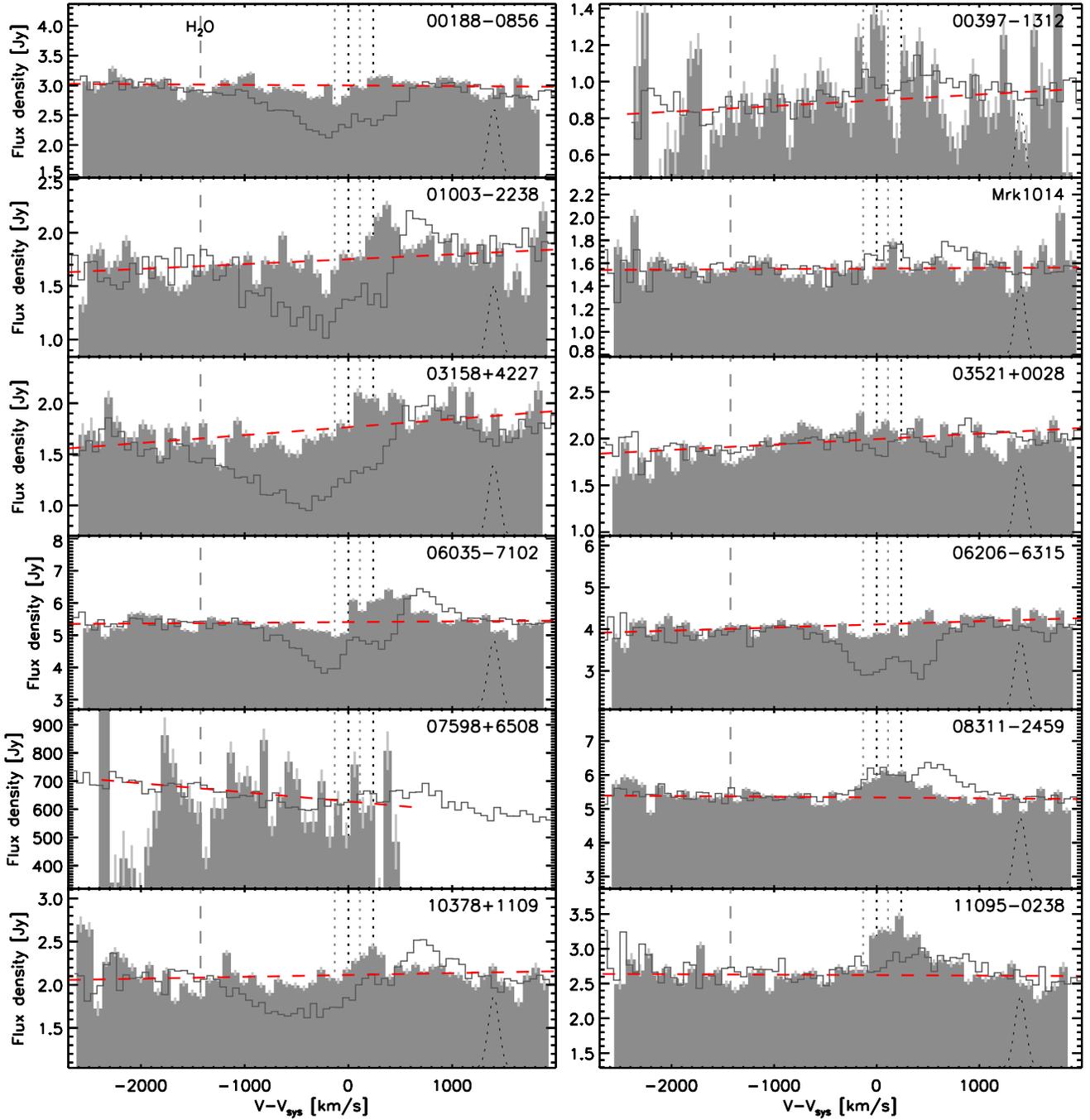}
\caption{Herschel-PACS spectra (grey surfaces) of the unresolved blend 
of the 79\,$\mu$m OH and $^{18}$OH doublets for ULIRGs observed in the 
HERUS survey. All spectra have the same flux dynamic range.
Systemic velocity is defined as the rest wavelength of the OH 
$\Pi_{1/2}$ -- $\Pi_{3/2}$ $\frac{1^-}{2}$ -- $\frac{3^+}{2}$
transition at 79.118\,$\mu$m. The rest velocities of the 79\,$\mu$m
$^{16}$OH and $^{18}$OH lines (see Table\,\ref{tbl-4}) are indicated 
by black and grey dotted vertical lines, respectively. Also indicated, 
as a grey dashed vertical line, is the rest velocity of the H$_2$O 
4$_{23}$--3$_{12}$ transition (78.742\,$\mu$m; at $v$=--1425\,km s$^{-1}$). 
The adopted local continuum is shown in red. Overlaid also are the 
normalized OH119 profiles (dark grey histogram lines), scaled to 
the 79\,$\mu$m wavelength range using the adopted 79\,$\mu$m
continuum. The spectral resolution at 79\,$\mu$m is shown by a 
dashed black gaussian profile in the right bottom corner of each panel. 
\label{oh79_spectra}}
\end{figure*}

\begin{figure*}
\addtocounter{figure}{-1}
\includegraphics[scale=1.0]{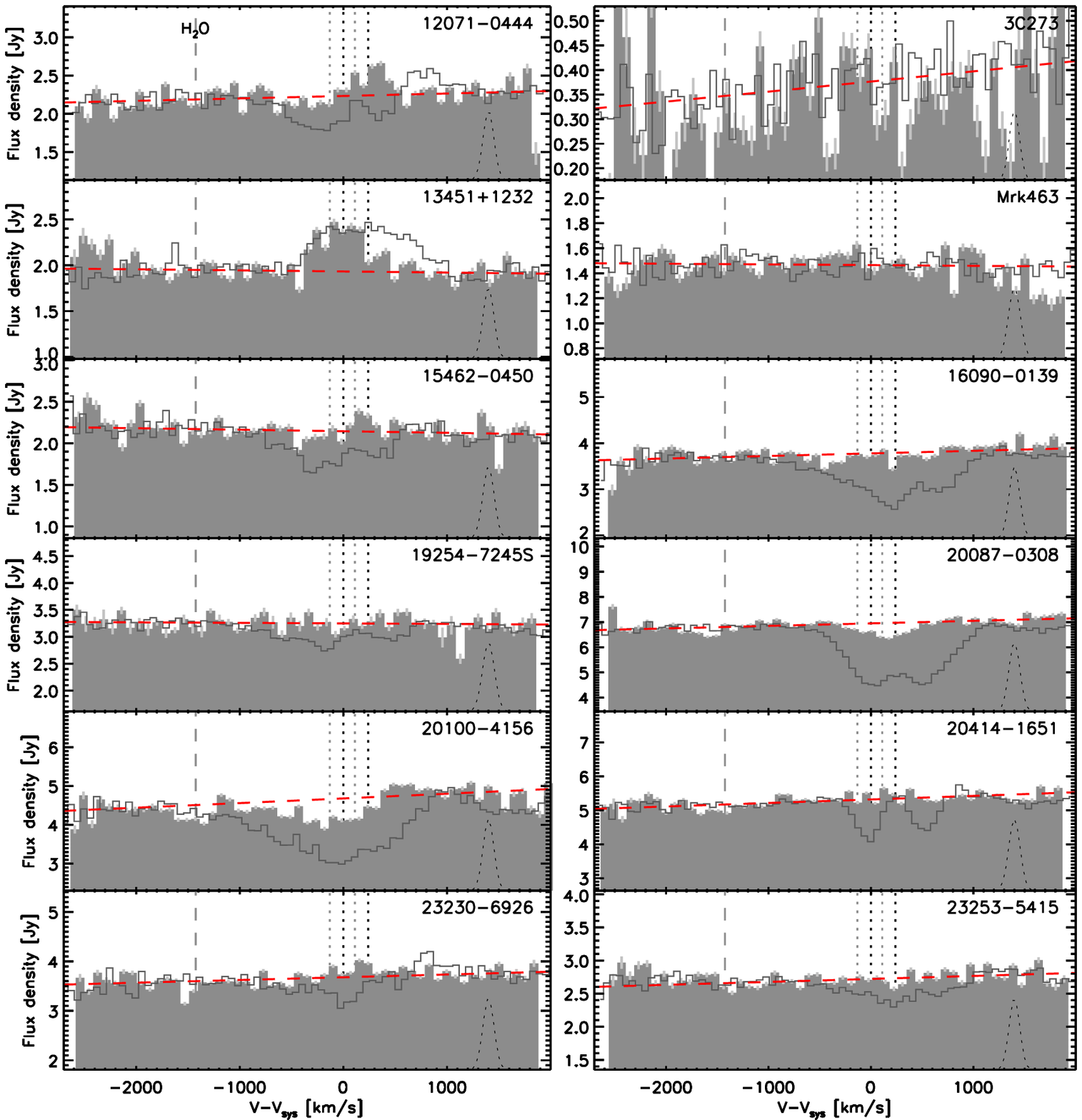}
\caption{79\,$\mu$m OH (continued).}
\end{figure*}


\begin{figure}
\includegraphics[scale=0.4]{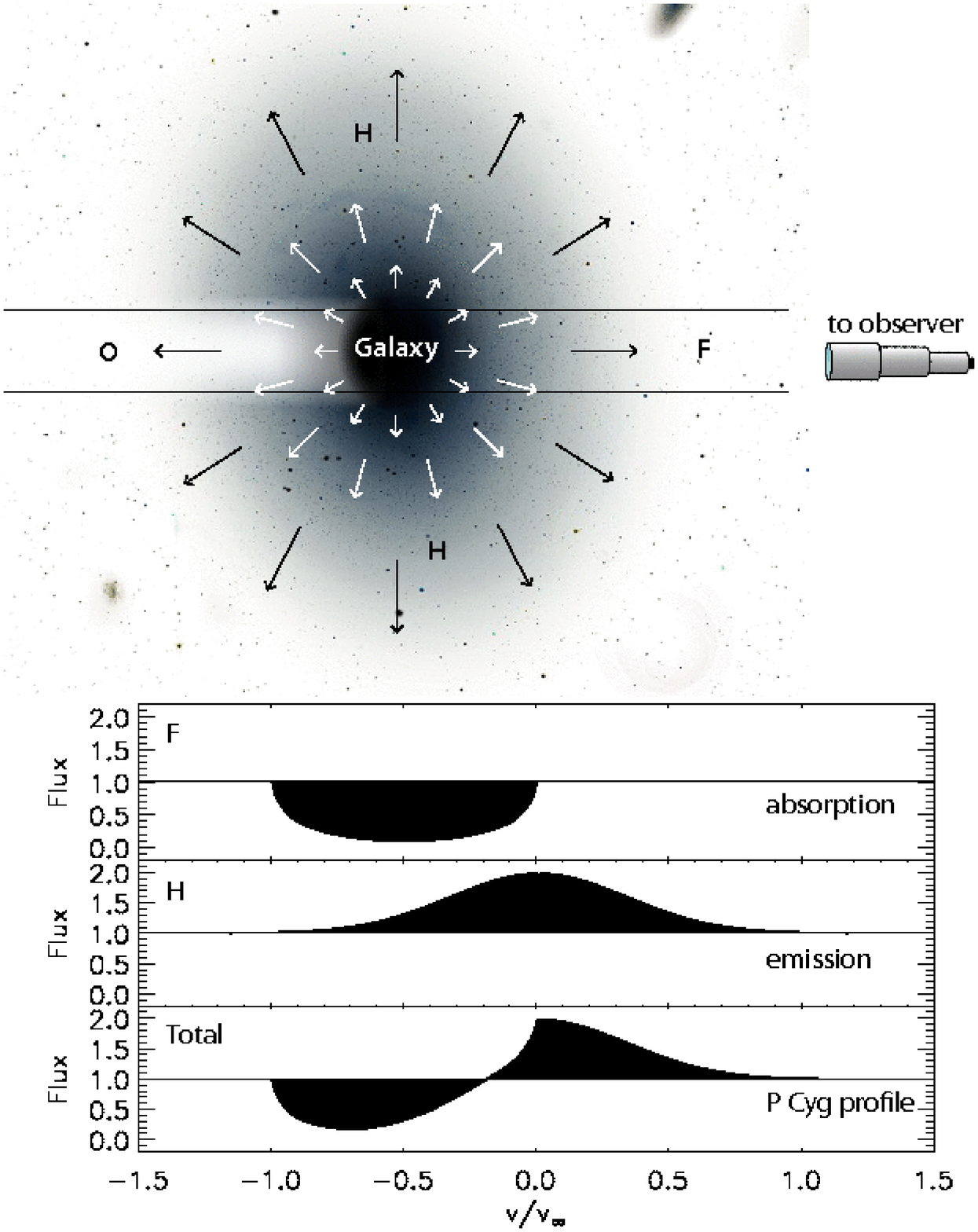}
\caption{Top panel: the geometry of a spherically symmetric
outflow with outward increasing velocity. We distinguish
four different components: the galactic nucleus, the 'tube'
in front of the nucleus (F), the occulted/attenuated 'tube'
behind the galaxy's nucleus (O), and the 'halo' around the 
nucleus (H).
Bottom panel:
A P Cygni profile arises from four spectral components:
gas in the front tube (F) absorbing the continuum spectrum 
of the galaxy's nucleus is coadded to the emission contributed 
by the halo (H) and any attenuated emission seeping through 
from the rear 'tube' (O). The profile depicted is for a single
line. In the case of OH119 and OH79 doublets two line profiles, 
520 and 240\,km s$^{-1}$ apart, respectively, need to be co-added.
This severely complicates the interpretation of the observed
spectral profile.
Figure adapted from \cite{lamers99}.}
\label{toymodel}
\end{figure}

\subsection{A qualitative analysis of the OH119 profiles}\label{oh119quali_section}

\subsubsection{An idealized outflow model}

P Cygni line profiles are powerful diagnostics of outflow kinematics.
Given the prevalence of P-Cygni type profiles among our sample
of OH119 spectra, we explore a general explanation 
for our observations along the lines of a spherically symmetric, 
optically thin, accelerating outflow. In our model we  
assume pure radiative excitation and de-excitation in a 2-level 
system, i.e. pure scattering, and no extinction. Effects of
extinction will be considered later.
In this idealized system (Fig.\,\ref{toymodel}) the absorption 
in the outflow arises in the 'tube' (F) in front of the 
far-IR continuum source (the nucleus), where radially outflowing 
OH molecules in the ground-state absorb radially emitted 119\,$\mu$m 
continuum photons. Like elsewhere in the outflow (which we will 
refer to as the 'halo' (H) from here on in) the absorption is 
followed by reemission\footnote{Instead of reemission of a 119\,$\mu$m photon 
the capture of an 84\,$\mu$m photon can excite the molecule 
to the next higher excited level in the $\Pi_{3/2}$ branch.
The chances of this occuring increase with 84\,$\mu$m photon
density, and thus with proximity to the far-IR nucleus.} 
of the photon. The chances of a reemitted 
continuum photon continuing along the same radial path 
are small, given the isotropic nature of radiative
deexcitation. The net observed result will hence be a blue shifted 
absorption trough with velocities ranging from zero to 
-$v_{max}$, the highest blue shifted velocity in the outflow.

All other parts of the outflow (the 'halo') contribute with
emission components to the profile, resulting from the isotropic 
reemission of 119\,$\mu$m photons which were absorbed on 
their radial paths by OH molecules in the outflow. 
The line-of-sight velocities sampled by the emission component 
range between +$v_{max}$ and -$v_{max}$, resulting in a symmetric
emission profile centered on zero velocity. 

In the absence of extinction and occultation, as per our
assumptions, none of the continuum photons from the central 
source will be lost. The photons are merely scattered into
or out of the line of sight to the observer. Since every
observer must see the same line profile under our assumption
of isotropy, this then implies that the number of photons 
scattered out of the line of sight must be equal to the number 
of photons scattered into the line of sight, or else some 
observers would see more emission than absorption, or vice versa. 
The net observed result of the interaction of 119\,$\mu$m continuum 
photons with outflowing OH molecules will thus only be a change 
in the velocity distribution of these photons between -$v_{max}$ 
and +$v_{max}$ 
\citep[see also][Gonz\'alez-Alfonso et al. in prep.]{gonzalez98}, 
since the total flux is effectively conserved.

\subsubsection{Problems with idealized outflow model}

In practice however, none of our OH119 P-cygni profiles display 
an emission component that is as strong as the absorption component.
In most sources the emission component is clearly weaker
than the absorption component. In these sources extinction
may suppress part of the 'halo' emission from reaching the
observer, especially the 'halo' component originating from
behind the dusty nucleus (labeled the 'occultation zone' ('O')
in Fig.\,\ref{toymodel}). This would 
especially suppress the most red shifted emission (close
to +$v_{max}$). The latter may be seen most clearly in 
IRAS\,01003--2238, where the blue absorption wing of the 
119.233\,$\mu$m OH line extends about 700\,km s$^{-1}$ beyond 
the red emission wing of the other (119.441\,$\mu$m) doublet 
line. Also absorption by CH$^+$ (at v=+1543\,km s$^{-1}$)
may contribute to the suppression (e.g. IRAS\,03158+4227).

\begin{figure*}
\begin{center}
\includegraphics[scale=0.75,angle=0]{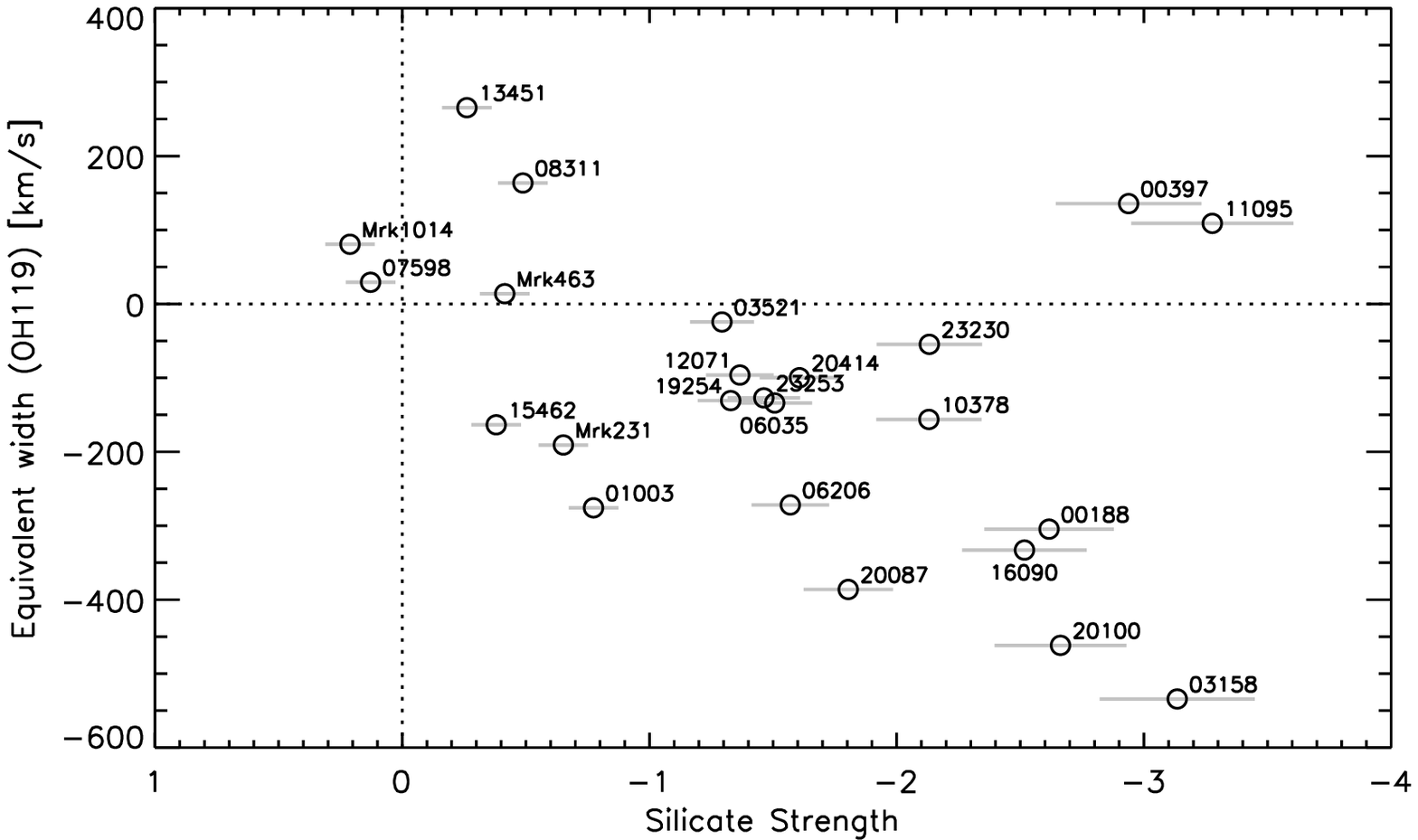}
\end{center}
\caption{The equivalent width of the OH119 feature as a 
function of silicate optical depth \citep{spoon07}. A negative 
equivalent width implies that the OH119 absorption component is 
stronger than the OH119 emission component. A positive
silicate strength implies the 9.8\,$\mu$m silicate feature to
be in emission. Our idealized model requires the equivalent 
width to be zero in the absence of extinction.
\label{tausil_intnormflux}} 
\end{figure*}

A close connection between the OH119 profile shape and 
obscuration is indeed implied by a comparison of the 
equivalent width of the OH119 profile, EQW(OH119), and the silicate 
strength S$_{sil}$ \citep{spoon07}. According to our outflow 
model EQW(OH119) should be zero in absence of extinction. 
We show this comparison in Fig.\,\ref{tausil_intnormflux} 
where we see a clear trend of decreasing OH119 emission 
relative to OH119 absorption as the silicate obscuration 
increases (S$_{sil}$ becomes more negative). This indicates that 
the obscuring geometry strongly affects the ability of the 
OH119 emission component to emerge, and that (at the very 
least for the most deeply obscured ULIRGs in our sample) 
the OH119 emission region is located within the buried nucleus, 
consistent with radiative excitation as the principal source 
of OH excitation. It also indicates that for the most 
strongly affected sources not only the 119\,$\mu$m continuum
but also the 79\,$\mu$m continuum will be optically thick.

Two sources in our sample are clear outliers
in Fig.\,\ref{tausil_intnormflux}: 
IRAS\,00397--1312 and 11095--0238. Both are deeply obscured
and only show OH119 emission components. Their OH119 pure
emission profiles, along with those of IRAS\,08311--2459, 
IRAS\,13451+1232 and Mrk1014, cannot be explained by our 
model.
In principle, in these sources the upper level of OH119 can 
be collisionally rather than radiatively excited to produce
the double emission profile. The emitting gas would hence 
have to reside in warm dense molecular clouds \citep{spinoglio05}.

There are many factors which likely condemn our outflow model to be 
an oversimplification of the true OH kinematics. For instance, the 
OH119 lines are mostly optically thick, meaning that they do not
give an accurate account of how the OH column
density is distributed over the various velocity components 
($^{18}$OH120 is optically thin). The outflow may also 
deviate drastically from spherical symmetry and may be confined
to outflow cones instead. This would introduce orientation 
and filling factor effects with consequences for both the observed 
line-of-sight velocity field and the relative strength of emission 
and absorption components. For instance, an outflow with a cone 
axis directed towards the observer would have a dominant 'tube' 
component with less 'halo' emission than in a spherically symmetric 
outflow (e.g. IRAS\,06206--6315). In addition, we would expect to 
see the full outflow velocity field. In practice it will
be hard to distinguish this scenario from obscuration effects 
discussed earlier. Similarly, an outflow directed in the plane 
of the sky would result in 'halo' emission that is strong compared
to the 'tube' component.  The OH gas would then have to 
display minimal line of sight velocities, consistent with a 
mainly tangential velocity field (e.g. IRAS 13451+1232). 

Even more complex scenarios may have to be invoked to explain
the existence of redshifted absorption in combination with
blue shifted absorption, like seen in IRAS\,16090--0139.
It is possible that the line-of-sight absorption traces a slow
inflow in combination with a fast outflow in this source. 

A question that remains is why only 5/24 sources in our sample
show pure OH119 emission profiles, two of which are sources
with strong silicate obscuration (IRAS\,00397--1312 and 
IRAS 11095--0238).

\subsection{Maximum OH119 outflow velocity ($v_{max}$) and balnicity}\label{vmax_section}

Theoretical studies of the ram pressure exerted by supernova-heated 
winds and radiation pressure exerted on dust grains by clusters of 
OB stars and AGN accretion disks predict the power of these wind 
acceleration mechanisms to crudely correlate with the maximum velocity 
of the outflowing gas. Determination of the maximum gas velocity in 
an outflow is hence of great diagnostic value.

Both the doublet nature of the OH119 line and the proximity of the 
$^{18}$OH120 doublet complicate a straightforward determination of the 
maximum outflow velocity ($v_{max}$) of the OH gas. The only OH119 doublet 
line from which we can measure $v_{max}$ without having to take into
account contamination by another doublet line is the 119.233\,$\mu$m line. 
And even for this line it is only the bluest 500\,km s$^{-1}$ that 
is free of contamination by the other (119.441\,$\mu$m) doublet 
line.

\begin{figure*}
\begin{center}
\includegraphics[scale=0.9,angle=0]{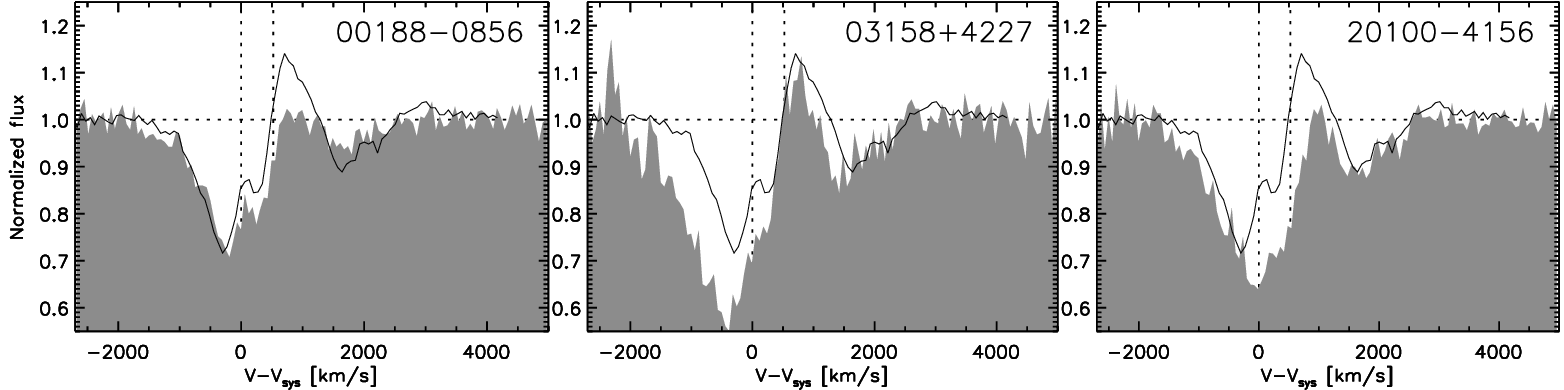}
\end{center}
\caption{Comparison of the normalized OH119 profiles of
IRAS\,00188--0856, 03158+4227 and 20100--4156 (grey surfaces) 
to the OH119 profile of Mrk231 \citep[{\it black} line;][]{fischer10}.
\label{mrk231_00188_03158_20100_comp}} 
\end{figure*}

Here we define the maximum outflow velocity $v_{max}$ as the
velocity at which a b-spline fit to the blue absorption wing of 
the normalized OH119 profile intercepts the continuum. This 
is a different definition than used by the SHINING team 
\citep{sturm11}, who obtain their value of $v_{max}$ as part of their 
model fit to several transitions between multiple levels of the OH 
molecule. Similar observations of transitions between excited OH
levels do not exist for the HERUS sources. For Mrk231 the results 
of the two methods agree within their uncertainties: 
$v_{max}$=1400$\pm$150\,km s$^{-1}$ using the SHINING method 
\citep{fischer10} and $v_{max}$=1573$\pm$200\,km s$^{-1}$ 
using ours. The B-spline fits for our sample
are shown in Fig.\,\ref{normalized_oh119_spectra} as green dashed
curves. The resulting maximum outflow velocities are tabulated in
Table\,\ref{tbl-2}. We estimate the uncertainty in our measured
maximum outflow velocity to be $\pm$200\,km s$^{-1}$, reflecting
the uncertainty in continuum placement, the signal-to-noise of
the spectrum, and the adopted value for the turbulent velocity 
(100\,km s$^{-1}$).
Three ULIRGs in our sample have maximum outflow velocities comparable 
to or exceeding that of Mrk\,231. The OH119 profiles of these three 
sources, IRAS 00188--0856, 03158+4227 and 20100--4156, are 
compared to the OH119 profile of Mrk231 in Fig.\,\ref{mrk231_00188_03158_20100_comp}.

Following \cite{weymann91} in concept, we further determine the 
'balnicity' for those sources in our sample with pronounced OH119 
absorption troughs. Here we define the balnicity as the integrated 
normalized flux F($v$) below the 98\% flux level at velocities more 
blue shifted than -200\,km s$^{-1}$. 

\begin{equation}
B ({\rm km\ s^{-1}}) = \int_{-v_{\rm max}}^{-200} [1-\frac{F(v)}{0.98}]\,C\,dv
\end{equation}

The constant C is zero except for the velocity range in which the
normalized flux F($v$) is lower than 0.98, where C is unity. The
balnicity is hence very similar to the equivalent width between
two indicated velocities. The 
rationale for contributions to the balnicity only to be counted
for velocities more blue shifted than -200\,km s$^{-1}$ is that
this excludes OH gas at rest. Also, our choice of 98\% of
the normalized continuum as the cut-off for integrating the
absorption profile is rooted in the uncertainty in our choice 
of the continuum level. Excluding the first 2\% of the absorption
feature therefore ensures that we are never integrating a 'false' 
absorption.
The resulting balnicities have an estimated uncertainty of 15\%
and are tabulated in Table\,\ref{tbl-2}. Five ULIRGs have 
balnicities comparable to or exceeding that of Mrk\,231.

\begin{deluxetable}{lrrr}
\tablewidth{0pt}
\tablecaption{Quantities measured from the OH119
absorption component\label{tbl-2}}
\tablehead{
\colhead{Galaxy} & \colhead{$v_{\rm max}^{obs}$\tablenotemark{a}} &
\colhead{$v_{\rm max}$\tablenotemark{b}} & \colhead{Balnicity} \\
\colhead{} & \colhead{(km s$^{-1}$)} & \colhead{(km s$^{-1}$)}  & \colhead{(km s$^{-1}$)}
}
\startdata
00188--0856 &  -1805 &-1781   & 118\\
00397--1312 &        &        &    \\
01003--2238 &  -1272 &-1238   & 194\\
Mrk1014     &        &        &    \\
03158+4227  &  -2065 &-2044   & 350\\
03521+0028  &  -232  &$>$-100 & $<$1\\
06035--7102 &  -1155 &-1117   &  81\\
06206--6315 &   -805 &-750    &  31\\
07598+6508  &  -1225 &-1190   &  15\\
08311--2459 &        &        &    \\
10378+1109  &  -1332 &-1300   & 149\\
11095--0238 &        &        &    \\
12071--0444 &  -967  &-922    &  53\\
3C273       &        &        &    \\
13451+1232  &        &        &    \\
Mrk463      &        &        &    \\
15462--0450 &  -972  &-927    &  61\\
16090--0139 &  -1451 &-1422   &  53\\
19254--7245 &  -1163 &-1126   &  39\\
20087--0308 &  -864  &-812    &  24\\
20100--4156 &  -1635 &-1609   & 177\\
20414--1651 &  -249  &$>$-100 &   2\\
23230--6926 &  -894  &-845    &  20\\
23253--5415 &  -712  &-650    &   8\\
Mrk231\tablenotemark{c} &  -1594 &-1573 & 120\\
\enddata
\tablecomments{Estimated uncertainty for all $v_{\rm max}$ velocities:
  $\pm$200\,km s$^{-1}$, and for the balnicity 15\%.}
\tablenotetext{a}{Terminal velocity as observed.}
\tablenotetext{b}{Terminal velocity corrected for turbulence (100\,km s$^{-1}$) and
  spectral resolution in quadrature.}
\tablenotetext{c}{SHINING source \citep{fischer10} processed using our methods.}
\end{deluxetable}


\subsection{Correlations of $v_{max}$ and balnicity with 
other observables}\label{vmax_balnicity_section}

Here we follow the SHINING team \citep{sturm11} and look for a 
correlation between the maximum OH119 outflow velocity and 
measures of the power of the AGN and the starburst. 
To infer the power of the AGN in a ULIRG we measure the slope of 
the mid-IR continuum between 15 and 30\,$\mu$m and use the zero
point calibration from \cite{veilleux09} to convert the relative
AGN contribution to L$_{\rm bol}$, $\alpha$, to the bolometric AGN 
luminosity L$_{{\rm AGN},bol}$. We estimate the uncertainty in $\alpha$ 
to be $\pm$0.15. 
L$_{\rm bol}$ itself is assumed to be 1.15$\times$L$_{\rm IR}$, which 
\cite{veilleux09} deem appropriate for ULIRGs.
The star formation rate (SFR) is inferred from the fraction of
L$_{\rm IR}$ that is not powered by the AGN, 1-$\alpha$,
and is defined as SFR=(1-$\alpha$)$\times$10$^{-10}$ L$_{\rm IR}$
\citep{sturm11}.

\begin{figure*}
\begin{center}
\includegraphics[scale=0.75,angle=0]{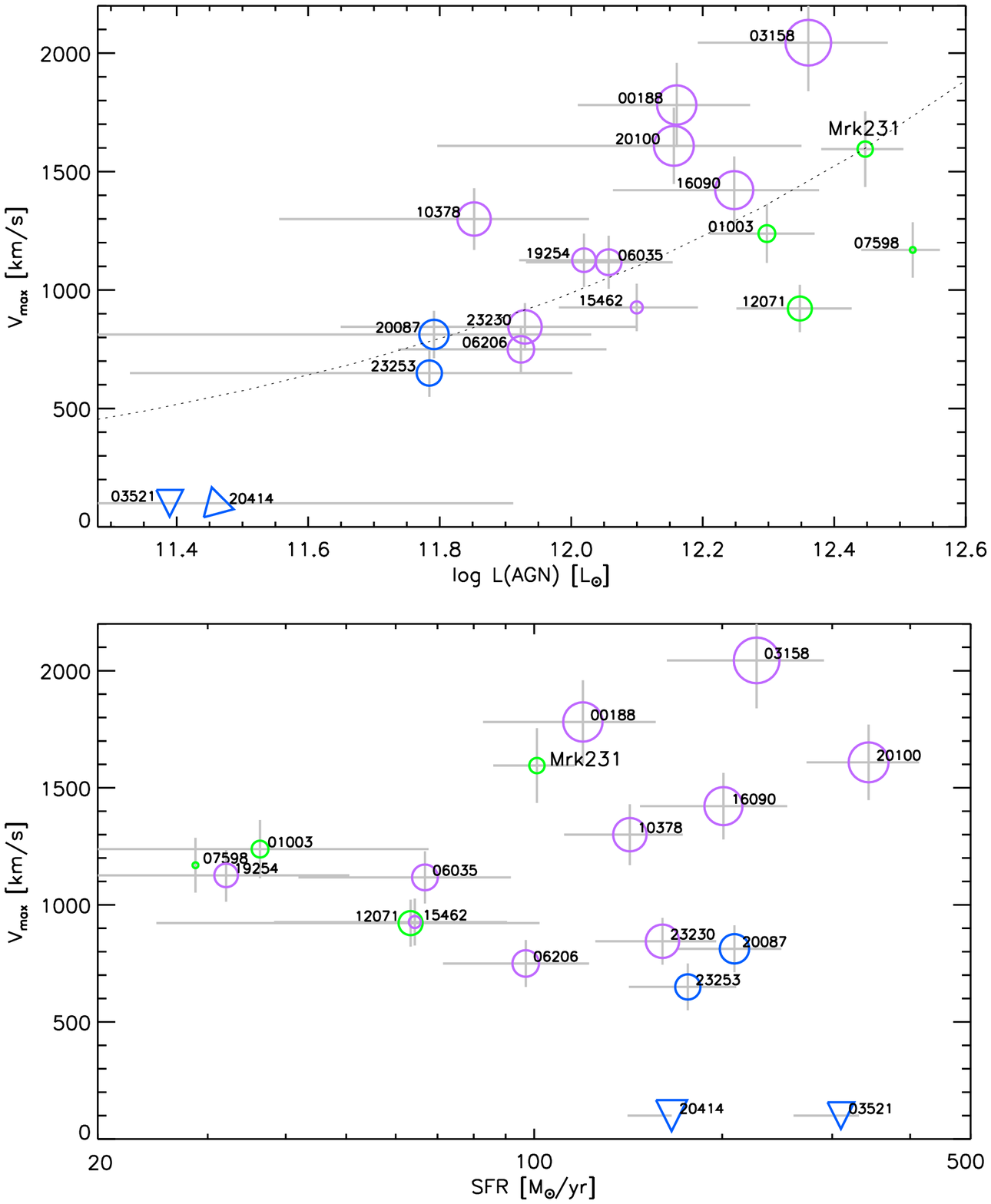}
\end{center}
\caption{Upper panel: Maximum OH119 outflow velocity as a function 
of AGN luminosity. The circle size is proportional to the silicate 
optical depth of the target. Triangles indicate upper limits.
The dotted line is a power law fit to all but two sources: 
IRAS\,03521 and 20414. The power law has the form
log(v$_{max}$/km s$^{-1}$)=-2.64($\pm$1.80)+0.47($\pm$0.15)$\times$log(L$_{{\rm AGN},bol}$/L$_{\odot}$).
Lower panel: Maximum OH119 outflow velocity as a 
function of star formation rate (SFR). In both panels
starburst-dominated sources ($\alpha$$<$0.25) are shown in 
blue, AGN-dominated sources ($\alpha$$>$0.75) in green, and
intermediate sources in purple.
\label{lagn_sfr_vmax}} 
\end{figure*}

In Fig.\,\ref{lagn_sfr_vmax} we plot the maximum outflow velocity
for 17 out of 24 of our sources as a function of AGN luminosity 
and star formation rate. Also shown is Mrk\,231, the SHINING source 
\citep{fischer10} which we analyzed in the same way as the
HERUS sample. The upper panel of Fig.\,\ref{lagn_sfr_vmax} 
suggests a trend of increasing maximum OH outflow velocity with 
increasing AGN power. The trend appears steeper for sources
with a high silicate optical depth than for sources with a 
lower silicate optical depth. All four starburst-dominated 
sources (color-coded blue) further show maximum outflow 
velocities below $\sim$800 km s$^{-1}$, while all four AGN-dominated 
sources (color-coded green) display maximum outflow velocities 
in excess of 900 km s$^{-1}$.
The lower panel of Fig.\,\ref{lagn_sfr_vmax} is a scatter plot,
suggesting no correlation to exist between the 
f(30\,$\mu$m)/f(15\,$\mu$m) based SFR and the maximum OH
outflow velocity in our sample. A similar lack of correlation
is obtained using a PAH flux based SFR \citep[][their Eq. 5]{farrah07}.
The above findings are consistent with the results presented by
\cite{sturm11} for a sample of 6 ULIRGs and one starburst galaxy
observed in the OH119, OH79 and OH65\footnote{Transition between the 
second and third excited states on the $\Pi_{3/2}$ ladder of OH. 
The doublet lines are centered at 65.132 \& 65.279\,$\mu$m, respectively.}
doublets.

\begin{figure*}
\begin{center}
\includegraphics[scale=0.75,angle=0]{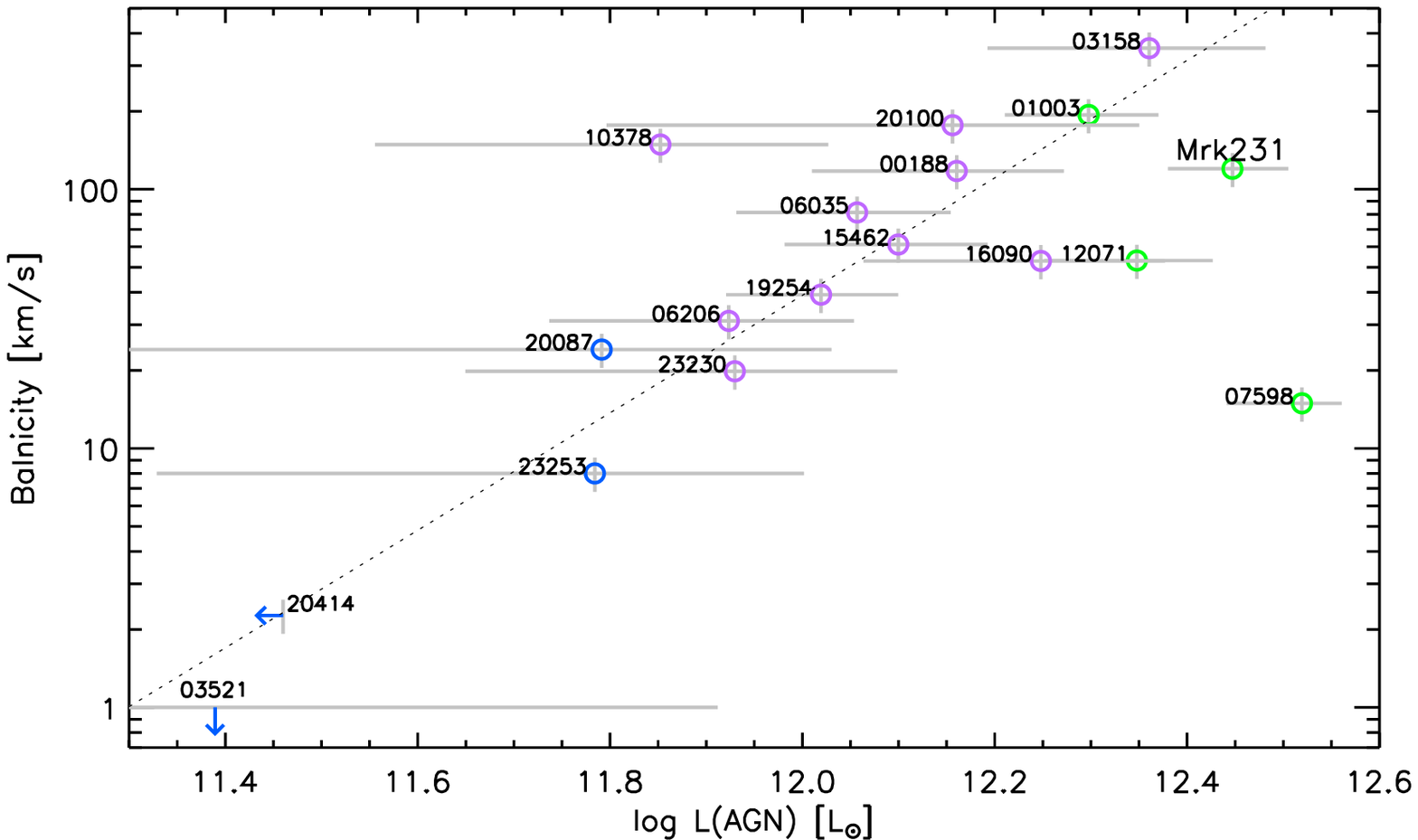}
\end{center}
\caption{The balnicity of the OH119 absorption profiles as
a function of AGN luminosity for our sources. Not shown are
sources with pure emission features. The dotted line is a power 
law fit to all but three sources: IRAS\,03521, 07598 and 20414.
The power law has the form 
log(B/km s$^{-1}$)=-25.6($\pm$9.6)+2.27($\pm$0.79)$\times$log(L$_{{\rm AGN},bol}$/L$_{\odot}$).
Starburst-dominated sources ($\alpha$$<$0.25) are shown in 
blue, AGN-dominated sources ($\alpha$$>$0.75) in green, and
intermediate sources in purple.
\label{lagn_balnicity}} 
\end{figure*}

As shown in Fig.\,\ref{lagn_balnicity}, a correlation also 
appears to exist between the AGN luminosity and the balnicity 
of the OH119 doublet. 
Apart from a few notable outliers the power law correlation holds 
over a factor 4 in AGN luminosity. The result may be 
surprising since the OH119 doublet is generally optically thick
and thus the depth of the feature is not necessarily a true 
measure of the OH column probed.
Outliers in this plot, IRAS\,10378+1109, 12071--0444, 16090--0139, 
Mrk231, and especially IRAS\,07598+6508, happen to be outliers also 
in the top panel of Fig.\,\ref{lagn_sfr_vmax}. Trends in both plots 
would improve if the AGN fraction $\alpha$ were to have been 
underestimated beyond our adopted uncertainty of 15\%  
for IRAS\,10378+1109 ($\alpha$=0.30), and overestimated 
for IRAS\,12071--0444 ($\alpha$=0.75), 16090--0139 
($\alpha$=0.43) and Mrk231 ($\alpha$=0.81). For IRAS\,07598+6508
no plausible change in AGN fraction can bring this source in
agreement with the general trend. Also an unfortunate choice for 
the local continuum offers no explanation for its shallow OH119
absorption component. Perhaps IRAS\,07598+6508 has 
more in common with the three high-luminosity ULIRGs lacking OH119 
absorption components, and which therefore are missing from the
above analyses: IRAS\,00397--1312, 08311--2459 and Mrk\,1014. Like
IRAS\,07598+6508, all three are clearly AGN-dominated based on their 
estimated AGN fractions (Table\,\ref{tbl-1}).

Note that based on balnicity, IRAS\,01003--2238, and not IRAS\,00188--0856, 
would be the second most extreme OH119 outflow source behind
IRAS\,03158+4227.

\begin{figure*}
\begin{center}
\begin{tabular}{c}
\includegraphics[scale=0.5,angle=90]{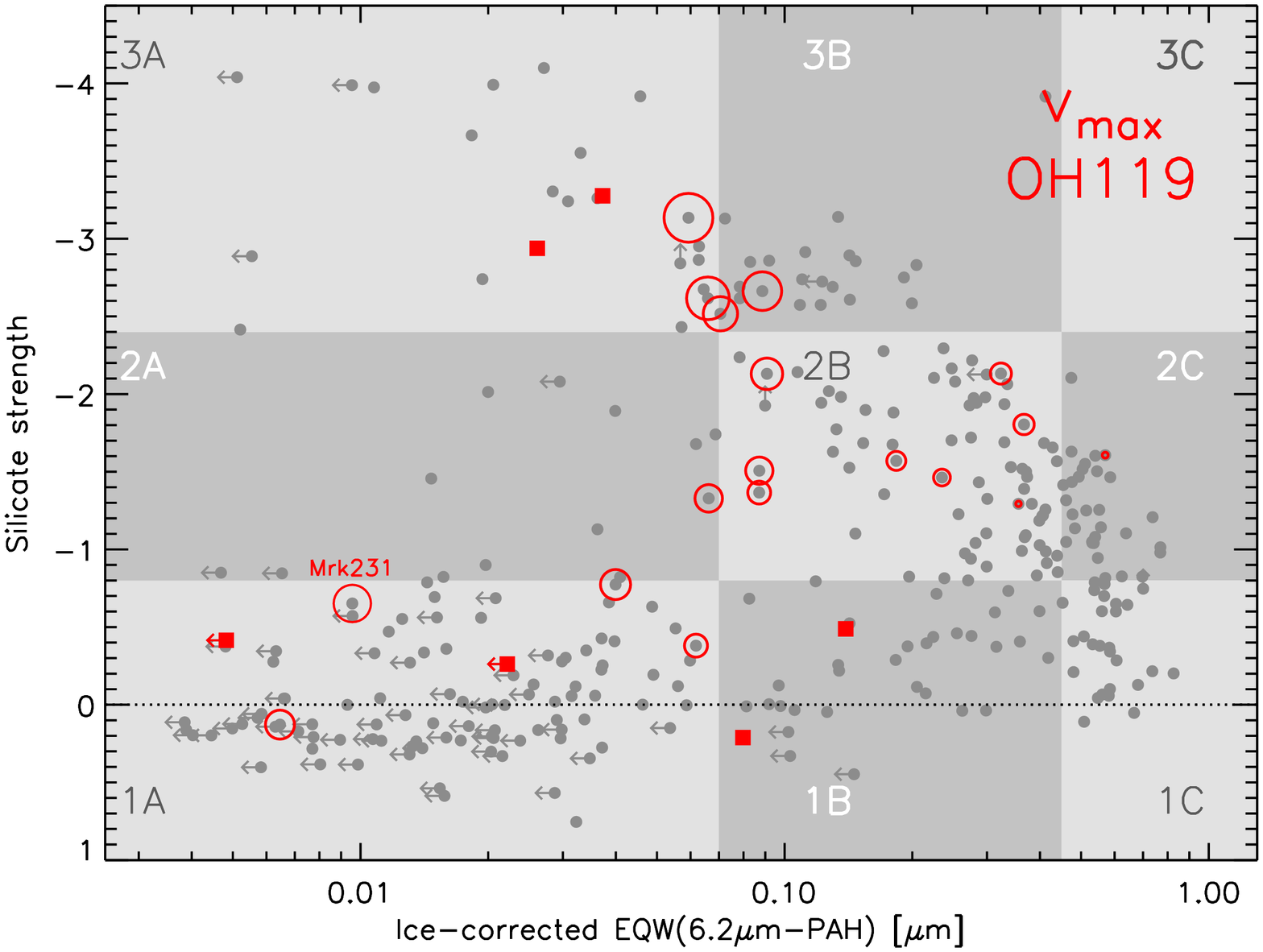} \\
\includegraphics[scale=0.5,angle=90]{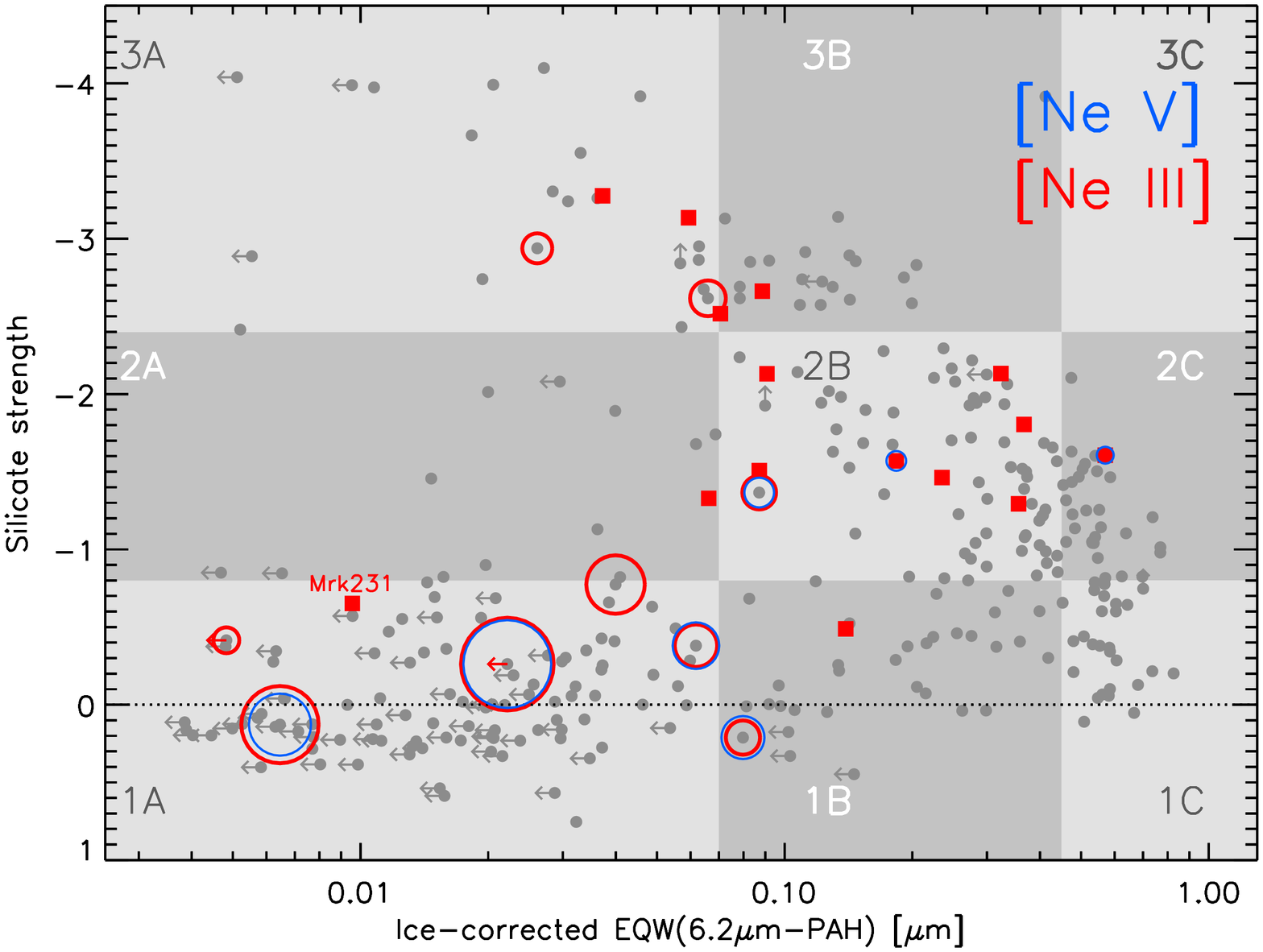}
\end{tabular}
\end{center}
\caption{Diagnostic plot of the equivalent width of the 6.2\,$\mu$m 
PAH emission feature versus the 9.7\,$\mu$m silicate strength \citep{spoon07}.
Upper and lower limits are denoted by arrows. The galaxy spectra 
are classified into 9 classes (identified by 9 shaded rectangles) 
based on their position in this plot. From class 1$\rightarrow$2$\rightarrow$3
the silicate absorption feature increases in strength, while from
class A$\rightarrow$B$\rightarrow$C the equivalent width of the
6.2\,$\mu$m PAH emission feature increases. 
Top panel:
circle sizes (red) are proportional to the maximum OH119 outflow 
velocity ($v_{max}$) inferred from the blue shifted absorption wing. 
Sources shown as red squares lack an OH119 absorption component 
from which to infer $v_{max}$. 
Bottom panel:
circle sizes are proportional to the blue wing velocity at which 
the neon line profile (15.56\,$\mu$m [Ne III]=red; 14.32\,$\mu$m [Ne V]=blue) 
reaches 10\% of the peak flux. Circles are plotted only for those
sources for which the line center shift at 10\% of the peak flux
(VC10) is more negative than -150 km s$^{-1}$. Note that [Ne V] 
lines generally are not detected below silicate strengths of -2.
\label{forkdiagram}}
\end{figure*}

In Fig.\,\ref{forkdiagram} we show the distribution of HERUS
sources in the mid-IR diagnostic diagram of 6.2\,$\mu$m PAH 
equivalent width (EQW62) versus silicate strength (S$_{sil}$),
the so-called 'fork diagram' \citep{spoon07}. This diagram
separates sources dominated by AGN clumpy hot dust emission (class 1A)
from sources dominated by star formation (class 1C) and sources 
dominated by a deeply buried nucleus (class 3A). Most sources 
are found along two branches connecting these three extremes. 
The 24 sources of the HERUS sample are distributed quite evenly
along these branches and are shown as red circles and squares. 
In the upper panel the circle size is proportional 
to the measured maximum OH119 outflow velocity ($v_{max}$), while 
red squares denote HERUS sources lacking an OH119 absorption component.
As can be seen in the upper panel, in our sample $v_{max}$ 
is highest among ULIRGs harboring a deeply buried nucleus,
and lowest among ULIRGs with a starburst-like mid-IR spectral
appearance.
HERUS ULIRGs in quadrant 1A (i.e. AGN-dominated sources) of the 
diagram show moderately high OH outflow velocities. Here and in 
quadrant 1B (which together constitute the AGN-dominated branch) 
we find a higher fraction of sources lacking an absorption 
component in their OH119 feature than on the obscuration-dominated 
branch (quadrants 2A+B and 3A+B).
In the bottom panel of Fig.\,\ref{forkdiagram} circle sizes 
are a measure of the highest blue shifted velocities of ionized 
gas seen in the emission lines of 15.56\,$\mu$m [Ne III] and 
14.32\,$\mu$m [Ne V] \citep{spoon09b}. The velocities are computed 
as the line center velocity shift at 10\% of the peak flux (VC10) 
minus half of the full width at 10\% of the peak flux (FW10).
The results are shown only for those sources in our sample
which display a genuine outflow wing. The criterion for the latter
is a VC10 line center blue shift of at least 150\,km s$^{-1}$
(see also Sect.\,\ref{mirfirline_section}).
As can be seen in this panel the highest outflow velocities
are found among the HERUS sources in quadrant 1A. 

\cite{spoon07} proposed that the strong silicate obscuration 
of ULIRGs on the upper branch of the fork diagram implies that 
these sources have not yet shed their obscuring molecular cocoon 
which was created during the earlier stages of the galaxy interaction.
In the fork diagram the evolutionary path that an energetically 
dominant buried AGN would take to evolve from the upper branch 
(quadrant 3A or 3B) to the locus of unobscured AGN (quadrant 1A)
would be almost vertical if the disruption/dispersion of the 
molecular cocoon by an AGN driven wind would expose the hot dust 
of the decloaking AGN. The additional strong mid-IR 
continuum emission would dilute the remaining starburst signatures 
and lower both the apparent silicate optical depth (i.e. increase 
S$_{sil}$) and the 6.2\,$\mu$m PAH equivalent width. The predominance 
of high {\it molecular} gas outflow velocities on the upper branch 
and the predominance of high {\it ionized} gas outflow velocities 
on the lower branch would be consistent with this evolutionary
scenario: the molecular cocoon needs to be sufficiently dispersed 
for the ionized gas outflow interior to it to become visible to us.

\subsection{Comparison to MIR-FIR fine-structure line profiles}\label{mirfirline_section}

\subsubsection{Mid-IR neon line profiles}

A survey of 82 galaxy-integrated Spitzer-IRS spectra of ULIRGs has
revealed that the emission line profiles of ionized neon gas show 
clear departures from virial gas motion in a sizeable fraction 
(28/82) of ULIRGs \citep{spoon09a,spoon09b}. 
The departures are both in the form of blue wings and blue shifts 
of the three strongest mid-IR neon lines, 12.81\,$\mu$m [Ne II], 
15.56\,$\mu$m [Ne III], and 14.32\,$\mu$m [Ne V]. This suggests 
decelerating ionized gas outflows in a stratified medium 
photo-ionized by the central source as the most likely 
explanation \citep{spoon09b}.
In Fig.\,\ref{normalized_oh119_spectra} we overlay the normalized 
neon line profiles onto the normalized OH119 profiles (in so far
as these lines are detected and have reasonable S/N) to facilitate
comparison of the kinematics of the molecular gas to that of the
ionized gas\footnote{The three neon lines together span a range of
21--97\,eV in ionization potential.}.
Note that the spectral resolution of the Spitzer-IRS observations 
is about half of that of the OH119 observations 
($\Delta$$v$ $\sim$500 and 250--300\,km s$^{-1}$, respectively). 

\subsubsection{[C II] line profiles}

\begin{figure}
\includegraphics[scale=1.0]{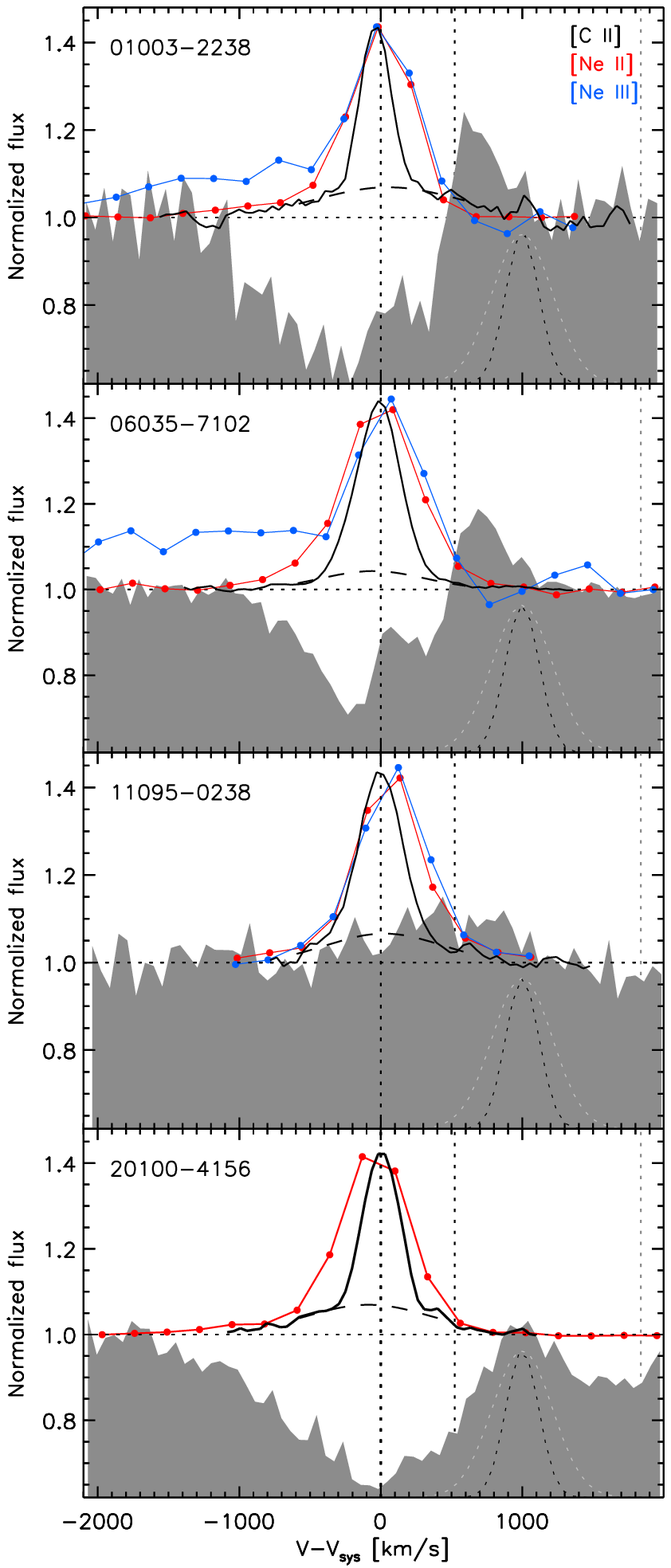}
\caption{Comparison of the normalized 119\,$\mu$m OH doublet 
profile to the scaled emission line profiles of 12.81\,$\mu$m 
[Ne II] (red), 15.56\,$\mu$m [Ne III] (blue) and 158\,$\mu$m 
[C II] (black). All four sources show broad [C II] pedestal 
emission, consistent with a gaussian profile of FWHM of 
800--1150\,km s$^{-1}$ (black dashed lines). The spectral 
resolution for the OH119 line is shown by a dashed black gaussian 
profile in the right bottom corner of each panel. The
Spitzer-IRS-SH spectral resolution is shown as a light 
grey profile. 
\label{c2_profiles}}
\end{figure}

Also overlaid in Fig.\,\ref{normalized_oh119_spectra} are the  
line profiles of the 158\,$\mu$m [C II] fine-structure line,
which are detected in all sources in our sample. Other FIR
fine-structure lines in our program are less frequently and less
well detected. The brightest of these other lines, 63\,$\mu$m \& 
145\,$\mu$m [O I] and 122\,$\mu$m [N II], do not show line shifts,
broad line wings or line asymmetries as far as we are able to 
assess at the limited S/N of these spectra. [C II], in contrast, 
does show departures from pure gaussian profile shapes. Some 
are in the form of small line center shifts of less than 100\,km s$^{-1}$
(IRAS\,00188--0856 and 01003--2238), others in the form of
subtle blue or red wing asymmetries (IRAS\,12071--0444 and 23253--5415,
respectively). Broad [C II] emission components are seen in
several sources. In Mrk\,463 about half the flux is associated
with a gaussian component with FWHM$_B$$\sim$600\,km s$^{-1}$.
In four other sources the broad component (B) is harder to discern, 
and has less flux associated with it (Fig.\,\ref{c2_profiles}).
These sources are IRAS\,01003--2238 (FWHM$_B$=1150\,km s$^{-1}$),
IRAS\,06035--7102 (FWHM$_B$=800\,km s$^{-1}$), IRAS 11095--0238
(FWHM$_B$=900\,km s$^{-1}$), and IRAS\,20100--4156 
(FWHM$_B$=1000\,km s$^{-1}$). See also Sect.\,\ref{c2massoutflow_section}. 
The broadest [C II] lines in our sample overall are seen in 
IRAS\,19254--7245 and IRAS\,20087--0308. Both are flat-topped, 
and have FWHMs of approximately 600\,km s$^{-1}$, and require two 
gaussian components for a good line profile fit.

\subsubsection{[N II] line profiles}

For some sources in our sample the 122\,$\mu$m [N II] line is broader 
than the [C II] line -- more than would be expected based on 
the difference in spectral resolving power between their observed 
wavelengths. An extreme case may be Mrk\,1014, for which the 
intrinsic FWHMs differ by a factor 2 
(Fig.\,\ref{normalized_oh119_spectra}). We speculate that these
differences may be due to differences in critical density
between the 122\,$\mu$m [N II] and the [C II] lines in ionized 
gas, in the sense that the [N II] line may be tracing denser gas 
deeper in the galaxy potential well than the [C II] line does.
Note that some of our 122\,$\mu$m [N II] profiles are affected by 
absorption/emission due to the high-lying 4$_{32}$-4$_{23}$
transition of o-H$_2$O at 121.7191\,$\mu$m, 430\,km s$^{-1}$ 
in the blue wing of the [N II] line 
\citep[][]{fischer10,gonzalez10}, although absorption by
HF at the same wavelength cannot be ruled out \citep{fischer10}.

\begin{figure}
\includegraphics[scale=1.0]{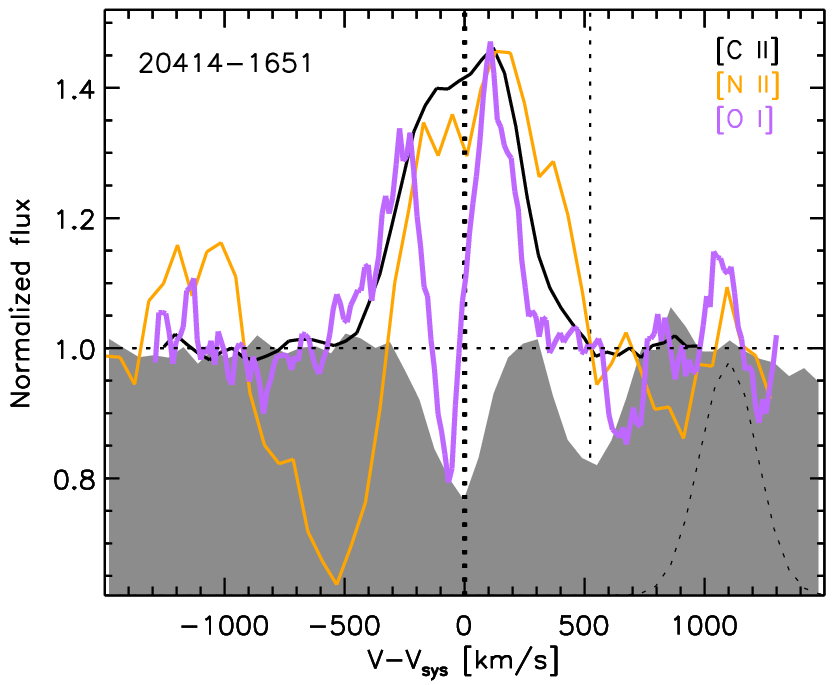}
\caption{Comparison of the normalized 119\,$\mu$m OH doublet profile 
to the scaled and smoothed emission line profiles of 63\,$\mu$m [O I] 
(purple), 122\,$\mu$m [N II] (orange), and 158\,$\mu$m [C II] (black). 
The strong absorption in the blue wing of the 122\,$\mu$m [N II] line 
is likely due to outflowing o-H$_2$O 4$_{32}$-4$_{23}$ 
\citep[121.7191\,$\mu$m; $v$=--440\,km s$^{-1}$;][]{fischer10}.
The [O I] line appears to be self-absorbed at velocities around 
-100\,km s$^{-1}$. Departures from pure gaussian behavior can
also be seen in [C II] and [N II] at those velocities.
The spectral resolution for the OH119 line is shown by a dashed 
black gaussian profile in the right bottom corner. 
\label{20414-profiles}}
\end{figure}

The clearest and most extreme 
example in our sample is IRAS\,20414--1651, whose [N II] line 
profile is shown in Fig.\,\ref{20414-profiles} along with 
the line profiles of [C II], 63\,$\mu$m [O I] and OH119. The
same source also shows the only clear case of self-absorption
of 63\,$\mu$m [O I] emission in our sample, indicating that
there is cold neutral oxygen gas (PDRs) in the line of sight 
to the central region \citep{poglitsch96}.
Another clear example of 121.7191\,$\mu$m H$_2$O 
absorption is IRAS\,20087--0308. H$_2$O {\it emission} is 
tentatively detected in IRAS\,00397--1312 at almost the same 
strength as the [N II] line there.

\subsubsection{Comparison of fine-structure line profiles
to OH119 profiles}

The source-by-source comparison of the galaxy-integrated 
fine-structure and OH119 line profiles as visualized in
Fig.\,\ref{normalized_oh119_spectra},\,\ref{c2_profiles} and
\ref{20414-profiles} allows us to make several observations
on the gas kinematics of the ionized, neutral and molecular
gas in our sample of ULIRGs as seen in the far-IR.
First, the far-IR fine-structure lines do not show the 
pronounced line asymmetries that the three mid-IR neon lines 
and the OH119 doublet show. Unlike the latter lines, the
far-IR fine-structure line emission is likely dominated
by low-velocity components which are closely tied to gas 
in virial motion in the galactic nuclei. 
Second, in 4/24 sources the far-IR fine-structure line with 
the highest S/N in our sample, [C II], shows a faint symmetric 
pedestal which can be fit with a gauss profile with 
FWHM$_B$=800--1150\,km s$^{-1}$. In three of these sources
the velocity field covered by the blue wing of the OH119 
absorption is in good agreement with the extent of the 
[C II] pedestal and of the 12.81\,$\mu$m [Ne II] emission line 
(Fig.\,\ref{c2_profiles}). Only in IRAS\,20100--4156 does the 
[C II] emission not extend to as high a blue shift as the 
OH119 absorption and the [Ne II] emission. Red wing [C II] 
emission appears to extend beyond the red wing of [Ne II] 
and the red wing of the red (119.441\,$\mu$m) doublet line 
of OH119 in IRAS\,01003--2238. This extension could point 
toward significant attenuation at wavelengths shorter than
158\,$\mu$m on the emission originating at the far side of
an outflow.
Third, the highest detected blue shifted velocities probed by 
the [C II] line are generally far smaller than those probed 
by OH119. There are, however, a few cases where the (modest) 
maximum blue shifted velocities match: IRAS\,13451+1232, 
20087--0308 and 23253--5415. 
Fourth, equally poor correspondence is found between the blue
shifted velocities probed by the three neon lines and OH119.
In 2/3 of our sample the neon lines probe higher blue shifted
velocities than OH119. Only in two sources it is the
other way around: IRAS\,03158+4227 and 20100--4156. Note,
however, the faint blue wing of [NeII] emission in the latter
source extends to almost the same maximum blue shift as
OH119 (Fig.\,\ref{c2_profiles}). Wings like these easily
remain hidden in lower S/N spectra.

The comparison of the velocity profiles of OH119 to the 
mid-IR and far-IR fine-structure lines hence illustrates 
the importance of OH119 and the mid-IR neon lines as tracers 
of molecular and ionized gas outflows in ULIRGs. Far-IR 
fine-structure lines, on the other hand, appear less suitable 
as outflow tracers, as they are much more strongly dominated 
by gas in virial motion than the neon lines and the OH119 
absorption profiles. Excellent S/N is hence required to be
able to reveal high-velocity wings in far-IR fine-structure 
lines. In our sample only 4/24 sources do that.

Interestingly,
the velocity field covered by the blue wing of the OH119 
absorption is in good agreement with the extent of the blue
wing of the [C II] and  12.81\,$\mu$m [Ne II] emission lines 
in three out of four sources (Fig.\,\ref{c2_profiles}).
Only in IRAS\,20100--4156 does the [C II] emission not extend
to as high a blue shift as the OH119 absorption and the [Ne II] 
emission.

\begin{figure*}
\begin{center}
\includegraphics[scale=0.75,angle=0]{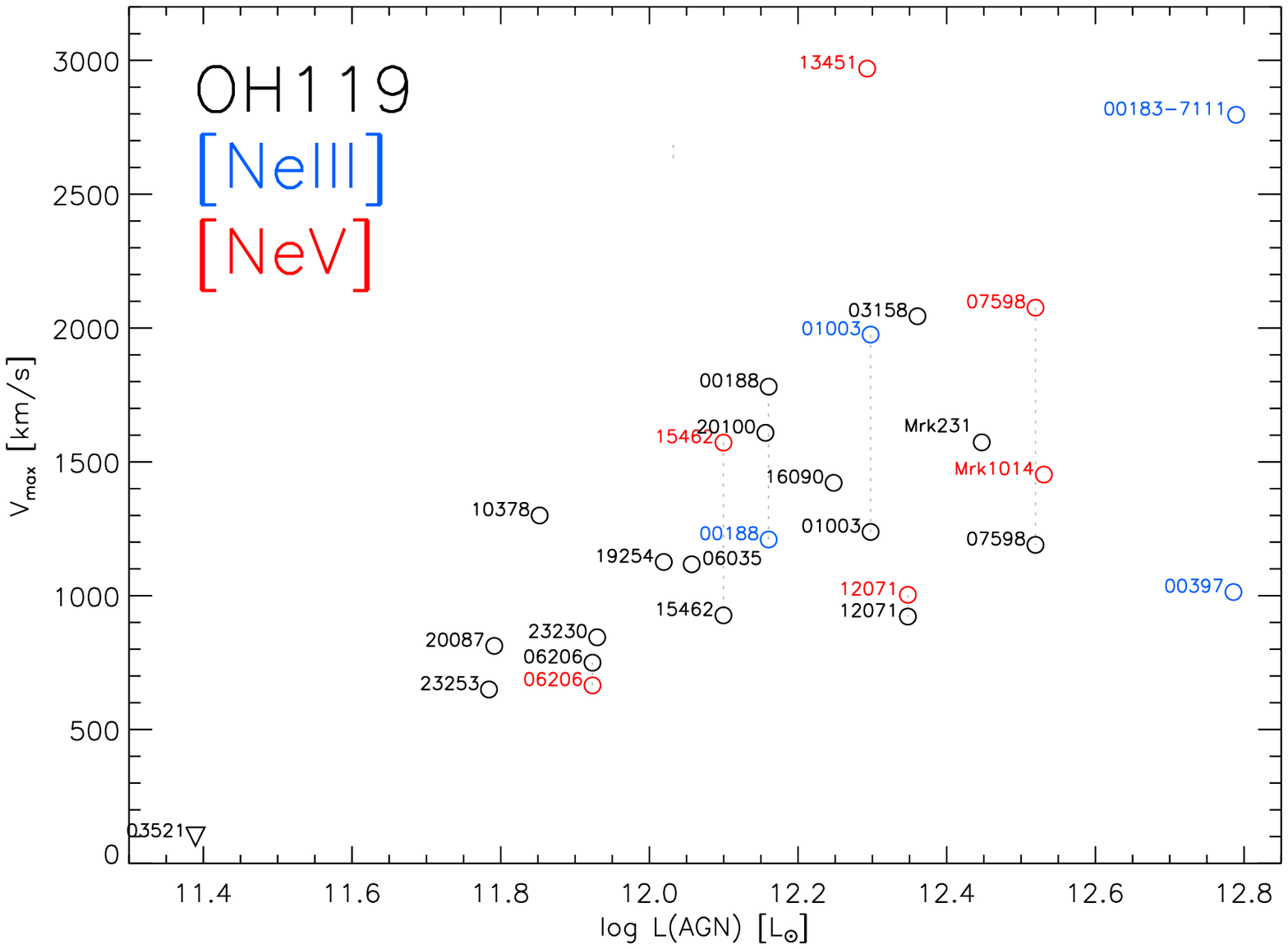}
\end{center}
\caption{Maximum outflow velocity as measured for OH119
absorption profiles (black), and 15.56\,$\mu$m [Ne III] (blue) 
and 14.32\,$\mu$m [Ne V] (red) emission line profiles. 
The OH119 data is the same as in Fig.\,\ref{lagn_sfr_vmax}. 
Besides the HERUS sample the plot contains the neon 
line measurement for IRAS\,F00183--7111 \citep{spoon09a}.
The triangle used for IRAS\,03521+0028 indicates an upper limit
to $v_{max}$(OH119). We estimate the uncertainty in $v_{max}$([Ne III]) 
\& $v_{max}$([Ne V]) to be $\sim$15\%. 
\label{lagn_oh_neon_vmax}} 
\end{figure*}

A different picture emerges if we define a maximum outflow velocity 
for the 15.56\,$\mu$m [Ne III] (blue) and 14.32\,$\mu$m [Ne V] lines
and compare their maximum outflow velocities to those of the OH119 data.
We define the maximum neon line outflow velocity ($v_{max}$(Ne)) as 
the velocity at which the blue wing of the emission line profile reaches 
10\% of the peak of the line profile. This quantity is defined
only for those line profiles which exhibit a clear blue wing or 
line profile shift at 10\% of the peak of the profile (VC10).
The comparison between the maximum outflow velocities for the 
molecular and ionized gas outflows is shown in Fig.\,\ref{lagn_oh_neon_vmax}. 

While neither the [Ne III] or the [Ne V] results show a correlation 
with AGN luminosity that is tighter than the correlation found for 
OH119 (Fig.\,\ref{lagn_sfr_vmax}), the three tracers combined do seem 
to map out a 'zone of avoidance' in Fig.\,\ref{lagn_oh_neon_vmax} in 
which only one of our sources is found:
IRAS\,13451+1232. Ignoring this source for the moment, 
the results suggest the existence of a maximum attainable 
maximum outflow velocity at a given AGN luminosity, and 
this maximum value to increase with increasing AGN luminosity.
At AGN luminosities below 10$^{12.35}$ L$_{\odot}$ the upper envelope
is sampled predominantly by the molecular outflow velocities, 
while above  10$^{12.35}$ L$_{\odot}$ it is the ionized gas (i.e. 
the [Ne III] lines of IRAS\,F00183--7111 \citep{spoon09a} 
and IRAS\,07598+6508) that probes this trend.
The existence of a maximum attainable outflow speed rather than 
a noisy correlation of $v_{max}$ and L$_{\rm AGN}$ may be a natural result 
of projection effects of the outflows due to the random orientation 
of the molecular and ionized gas outflows within our sample.

The source 'violating' the zone of avoidance in Fig.\,\ref{lagn_oh_neon_vmax}, 
IRAS\,13451+1232, 
cannot be reconciled with the trend by increasing the fraction 
of its L(IR) powered by the AGN, as this fraction is already very 
high ($\alpha$=0.82). While this may completely disprove the
concept of a maximum attainable outflow speed at a given AGN 
luminosity, IRAS\,13451+1232, a Gigahertz Peaked Spectrum (GPS) 
radio source, is known to have sub-kpc size radio jets \citep{holt03},
which may contribute to driving the ionized gas outflow.

\subsection{Atomic mass in outflow from [CII] pedestals}\label{c2massoutflow_section}

Four sources in our sample show high-velocity components 
in their [C II] lines (Fig.\,\ref{c2_profiles}) which can be
fitted with gaussian profiles of FWHM$_B$=800--1150\,km s$^{-1}$. 
These values are at the high end of gas velocities found for 
mid-IR AGN narrow line region (NLR) tracers 14.32\,$\mu$m [NeV] 
and 25.89\,$\mu$m [O IV] in ULIRGs 
\citep[e.g. IRAS\,05189--2524 and Mrk\,1014;][]{dasyra11}.
The high-velocity [C II] emission may hence originate from 
NLR gas that is sufficiently shielded to be only singly ionized.

Alternatively, and more likely, the high-velocity [C II] 
emission originates in outflow, as previously suggested to
explain the high-velocity wings in the [C II] profile of Mrk231 
\citep{fischer10} and the high redshift quasar 
SDSS\,J114816.64+525150.3 \citep{maiolino12}. Following
\cite{hailey10} and \cite{maiolino12} (their equations 1) 
we use the [C II] line luminosities of the broad component
to infer a {\it lower} limit to the total atomic mass in outflow. 
We find values ranging from 2$\times$10$^8$ M$_{\odot}$ to
4.5$\times$10$^8$ M$_{\odot}$ (IRAS\,20100--4156), which is
of the same order as the molecular gas mass seen in outflow
in Mrk231 (5.8$\times$10$^8$ M$_{\odot}$) using CO(1-0)
\citep{feruglio10}.

\section{Discussion and conclusions}

We have analyzed the Herschel-PACS spectra of the two lowest
ground-state transitions of OH and $^{18}$OH in a sample of
24 ULIRGs. The OH119 and OH79 doublets reveal P Cygni type 
profiles, and thus molecular outflows, to be common among 
late stage mergers. Smaller subsets of sources in our sample
either show pure absorption or pure emission profiles. While 
these do not necessarily imply outflows, plausible outflow 
geometries exist which are consistent with their observed 
line profiles. These geometries require the outflows to be 
highly obscured (to suppress the emission component) or 
highly non-spherical in nature (to explain pure emission 
or pure absorption profiles).

The molecular gas velocities probed by our observations span
a wide range, with some sources showing highest outflow 
velocities of no more than 100\,km s$^{-1}$ while others
show maximum outflow velocities of up to 2000\,km s$^{-1}$
in their absorption components. 
Absorption line studies of neutral gas outflows in ULIRGs 
\citep{rupke05a,martin05} suggests that starbursts drive 
outflows with highest outflow velocities of 500--700\,km s$^{-1}$. 
In our ULIRG sample 15/24 sources show maximum OH outflow 
velocities exceeding that, and 10/24 sources with maximum 
OH outflow velocities above 1000\,km s$^{-1}$. The large
majority of our sample ($\sim$2/3) would hence host 
AGN-driven molecular outflows, powered by radiation 
pressure on dust grains \citep{martin05,murray05,murray11}.

Correlation of both the maximum OH119 outflow speed and the 
balnicity (similar to the equivalent width of the blue shifted
absorption component; Eq. 1) to the star formation rate and 
bolometric AGN luminosity also identify the AGN as the 
likely driver of these molecular 
outflows. Since the balnicity is a reasonable proxy for the total 
momentum in the outflow, the AGN luminosity may correlate with 
the energy in the outflow.
These results confirm the findings by \cite{sturm11} for a 
smaller subset of local starburst galaxies and ULIRGs based 
on modeling of the observed OH119, OH79 and OH65 doublets.

Correlation with two other parameters, the 9.8\,$\mu$m silicate strength
and the equivalent width of the 6.2\,$\mu$m PAH emission
feature (Fig.\,\ref{forkdiagram}), suggests that the maximum 
outflow speed is generally
higher among moderately to deeply embedded ULIRGs than among 
ULIRGs with a starburst or AGN-like mid-IR spectral appearance.
This would suggest the most powerful molecular outflows to be 
associated with early stages of AGN-feedback, when the AGN is 
still deeply embedded. Two deeply obscured ULIRGs in our sample,
IRAS\,00397--1312 and 11095--0238 appear to be at odds with
this interpretation. Neither source shows a P-Cygni type 
high-velocity OH gas outflow in its spectrum. Instead both
show relatively narrow emission profiles. Both are also extreme 
outliers in a plot of the OH119 equivalent width as a function 
of silicate obscuration (Fig.\,\ref{tausil_intnormflux}) by 
showing OH119 purely in emission.
As proposed in 
Sect.\,\ref{oh119quali_section}, this spectral appearance 
may merely be an orientation effect resulting from the outflow 
being concentrated in a direction close to the plane of the sky.
IRAS\,00397--1312 and 11095--0238 may thus host similarly 
strong molecular outflows as the other deeply obscured
sources on the upper branch of the fork diagram 
(Fig.\,\ref{forkdiagram}). Furthermore, like
the other sources on that branch they may be evolving downward
towards the AGN locus (quadrant 1A in Fig.\,\ref{forkdiagram}) 
as the molecular outflow
continues to disrupt and dismantle the obscuring cocoon and 
expose more of the hot continuum associated with the AGN torus. 
At the implied high mass loss rates of hundreds to thousands of 
solar masses per year, as found by the SHINING team \citep{sturm11}, 
the timescale for depletion of the molecular gas reservoir and
transition to the AGN locus would range between 10$^6$ to 10$^8$ 
years. This would make this downward evolution in the fork
diagram a quick transition with relatively few sources caught
in mid-transit.

We have further compared the kinematic properties of the mid and 
far-IR fine-structure {\it emission} lines to the kinematic 
properties of the OH119 {\it absorption} components. 
Such a comparison is not straightforward, as the OH119 absorption
component probes the velocity field in the line of sight to the 
far-IR nucleus, and the emission lines trace gas all around the source. 
Comparison of the kinematic information gained from emission 
and absorption tracers may hence be subject to numerous biases
\citep[see e.g.][]{westmoquette12}.
The emission lines are density weighted, while the OH119 
doublet will be optically thick at most velocities and may not 
cover the entire far-IR continuum source. 
With these important caveats in mind, it may be no surprise 
that we find no correlation between the kinematic properties 
of the molecular and ionized gas on a source-by-source basis. 
And there may not be one, if the different gas phases in ULIRG
winds are not strongly coupled \citep{rupke11}.
Only when the maximum outflow velocities
of the OH and ionized neon gas are plotted together as a function 
of AGN luminosity the data suggests the existence of a maximum 
attainable outflow velocity for these species, with in some sources 
the ionized neon gas and in other sources the molecular gas reaching
this highest attainable velocity for a given AGN luminosity.
Whether or not this is a significant result may be tested by 
including neon and OH line profile measurements for other ULIRGs 
observed by Spitzer-IRS and Herschel-PACS as part of other 
programs. The latter results are, however, not yet published.
Further observations and further analysis will also have to
show whether there is a kinematic connection between the
decelerating ionized gas outflows, as implied by the mid-IR
neon line observations \citep{spoon09b}, and the molecular
gas outflows as traced by OH.

Finally, four of our sources show high-velocity wings in
their [C II] fine-structure line profiles which can be
fit with gaussian profiles with full width half maxima
ranging from 800 to 1150\,km s$^{-1}$. If interpreted as 
outflow signatures these [C II] pedestals imply maximum
outflow velocities up to 1000\,km s$^{-1}$ and neutral gas 
outflow masses of at least 2--4.5$\times$10$^8$ M$_{\odot}$. 
Given the lack of reasonable spatial constraints on both
the [C II] emitting and the OH absorbing and emitting 
envelopes we cannot infer mass outflow rates.

\acknowledgments

We are grateful to the SHINING team for their collegiality and
for providing us with the OH119 line profile of Mrk231. We further
would like to thank Jackie Fischer, Henny Lamers, Thomas Nikola and
Gordon Stacey for discussions, and Dan Weedman and Sylvain Veilleux
for sharing OH line scans prior to their publication. 
V.L is supported by a CEA/Marie Curie Eurotalents fellowship.
E.G-A is a Research Associate at the Harvard-Smithsonian
Center for Astrophysics, and thanks the support by the Spanish 
Ministerio de Econom\'{\i}a y Competitividad under projects
AYA2010-21697-C05-0 and FIS2012-39162-C06-01.
J.A acknowledges support from the Science and Technology Foundation 
(FCT, Portugal) through the research grants PTDC/CTE-AST/105287/2008, 
PEst-OE/FIS/UI2751/2011 and PTDC/FIS-AST/2194/2012. M.E would like to 
thank the support from ASTROMADRID through grant S2009ESP-1496, from 
Spanish MINECO: AYA2009-07304 and from ASTROMOL: CSD2009-00038.
Herschel is an ESA space observatory with science instruments provided 
by European-led Principal Investigator consortia and with important 
participation from NASA. Support for this work was provided by NASA 
through an award issued by JPL/Caltech.

{\it Facilities:} \facility{Herschel(PACS)}.

\end{document}